\documentclass[letterpaper, journal]{IEEEtran}
\usepackage{epsfig,graphics,subfigure,psfrag,float}
\usepackage{latexsym,amssymb,epsfig,subfigure,amsmath}
\usepackage{algorithm}
\usepackage{algpseudocode}
{

\newtheorem{Lem}{Lemma}

\newtheorem{Def}{Definition}

\newtheorem{Problem}{Problem}
\renewcommand{\QED}{\QEDopen}}

\newcounter{MYtempeqncnt}

\begin{document}
\title{Cross-Layer Design of FDD-OFDM Systems based on ACK/NAK Feedbacks}
\author{Zuleita~K.~M.~Ho, ~\IEEEmembership{Member,~IEEE,} Vincent
K.~N.~Lau,~\IEEEmembership{Senior~Member,~IEEE,} Roger S.~K.~Cheng
~\IEEEmembership{Member,~IEEE}\\}\maketitle
\markboth{
IEEE Transactions on Information Theory}{Cross-Layer Design of FDD-OFDM Systems
based on ACK/NAK Feedbacks}
\begin{abstract}
It is well-known that cross-layer scheduling which adapts power, rate and user
allocation can achieve significant gain on system capacity.  However,
conventional cross-layer designs all require channel state information at the
base station (CSIT) which is difficult to obtain in practice. In this paper, we
focus on cross-layer resource optimization based on ACK/NAK feedback
flows in OFDM systems without explicit CSIT. While the problem can be modeled
as Markov Decision Process (MDP), brute force approach by policy iteration or
value iteration cannot lead to any viable solution. Thus, we derive a simple
closed-form solution for the MDP cross-layer problem, which is asymptotically
optimal for sufficiently small target packet error rate (PER). The proposed
solution also has low complexity and is suitable for realtime implementation. It
is also shown to achieve significant performance gain compared with systems that
do not utilize the ACK/NAK feedbacks for cross-layer designs or cross-layer
systems that utilize very unreliable CSIT for adaptation with mismatch in CSIT
error statistics. Asymptotic analysis is also provided to obtain useful design
insights. 
\end{abstract}

\begin{keywords}
ACK, Acknowledgement, Cross-Layer, Feedback, Scheduling,   
Markov Decision Process, MDP, No CSI, Power Adaptation, Rate Adaptation
\end{keywords}
\section{Introduction}
\subsection{Background and motivation}
Cross-layer scheduling has been shown to achieve a significant
performance gain in wireless systems as a result of multiuser
diversity gain. Most of the existing cross-layer designs heavily
rely on either perfect CSIT \cite{Mecklenbrauker05}
\cite{Evans05}\cite{Liu05} or imperfect \cite{Vazquez05}
\cite{Lagunas04}/ delayed CSIT \cite{Jiang06} \cite{Du05}. 

\subsubsection{Absence of Accurate CSIT and CSIT error statistics}
Perfect
CSIT is difficult to obtain in practice, especially in FDD systems
in which explicit feedback is required. With imperfect CSIT
\footnote{There are two meanings behind "imperfect CSIT" in the
literature. The first meaning of imperfect CSIT refers to partial
knowledge of CSIT such as limited feedback but the partial CSIT
knowledge is received accurately (without errors) or timely (no
delay). On the other hand, the second meaning of imperfect CSIT refers
to inaccurate knowledge of CSIT (either with CSIT errors or
outdatedness). In this paper, the term "imperfect CSIT" refers to
the second meaning.}, systematic packet errors would result
even if powerful error correction codes are applied. This is
because given the imperfect CSIT, there is uncertainty on the
instantaneous mutual information at the base station and the scheduled data rate
may exceed the instantaneous mutual information, leading to packet errors
(channel outage) despite the use of powerful error correction coding.  It has
been shown \cite{Cioffi02}\cite{Leung06}
that packet errors cause significant degradation in cross-layer
performance. There are some works to take into account of
the imperfect CSIT or limited CSIT feedback in
cross-layer design. For example, in \cite{Tang05} \cite{Haleem05}, the authors
studied the cross-layer
design with noiseless limited feedback. In \cite{Vazquez05} \cite{Du05},
the authors studied OFDMA cross layer design with outdated CSIT. However, in all
these works, the CSIT obtained is either noiseless (or no delay) or the
statistics of the CSIT errors is assumed to be known \cite{Jiang06}. However, in
practice, the knowledge of CSIT errors statistics such as
CSIT error
variance and CSIT delay is
needed and this
is not easy to obtain because it depends on
the mobility of the users as well as the multipath profile. It is quite
challenging to have a robust cross-layer scheduling solution without the
knowledge of CSIT error variance. On the other hand, regardless of the CSIT,
there are always ACK/NAK flows between the mobiles (MS) and the basestations
(BS). A robust cross-layer scheduling should make the best use of the
ACK/NAK information which is embedded in the protocol. \footnote{in a similar
way as
\emph{outerloop power control} in CDMA
systems.} 

\subsubsection{Accomodation of mobiles with different receiver capability}
Conventional cross-layer design that utilized CSIT to perform resource
allocation is essentially an open-loop system because BS
cannot determine if the packet is received correctly or not even with the
knowledge of CSIT (due to decoding errors). In practice, the system
may have heterogeneous mix of mobiles with different capabilities (e.g. some has
turbo decoding capability while some only has simple detection capability). To
accommodate the heterogeneous mixture of receiver capability in the resource
allocation, the BS has to rely on ACK/NAK flows (because the ACK/NAK flows give
information about whether a packet can be decoded successfully or not). This
closed loop information cannot be obtained in CSIT-based
scheduler. 

\subsubsection{Heuristic Approach in existing literature}
Recognizing the importance of utilizing the ACK/NAK in the resource allocation
at the BS, there are existing works that discuss power control using
ACK/NAK feedbacks. However, most of the works either considered power control on
a wireless link only as well as utilizing heuristic algorithms or study the
performance by simulation. For example, a power adaptation design and
performance study utilizing ACK/NAK
feedbacks for point-to-point systems have appeared in
\cite{Park06}-\cite{Kim01}. Cross-layer scheduling utilizing
ACK/NAK feedbacks was investigated in \cite{Zuleita06}
\cite{Zuleita07}\cite{Chandramouli06}. In particular, power
control, rate adaptation and user scheduling for flat fading
channels and frequency selective channels were carried out in
\cite{Zuleita06} and \cite{Zuleita07} respectively whereas a rate
adaptation scheme based on ACK/NAK feedbacks was proposed in
\cite{Chandramouli06}. The authors proposed a 2-level hierarchy
stochastic scheduling algorithm based on \emph{learning automata}
(LA) for an AWGN channel  by rate adaptation.
Although the algorithm was shown to converge to the true channel
state values, the convergence is not proven to maximize the
throughput which is of usual practical concern. Moreover, in all
these works \cite{Zuleita06}\cite{Zuleita07}\cite{Chandramouli06},
the algorithm designs are based on heuristic solutions and it is
not clear what the best possible performance from the ACK/NAK
information is. Furthermore, the suboptimal solutions obtained
have high complexity and is not suitable for real-time
implementations. Moreover, in all these existing designs, there is
no mechanism to control the per-user packet error rate PER to a given target
level.
Yet, being able to control the PER of the wireless sessions per
user is very important from the requirements of applications
(e.g. voice and video codec).

Motivated by all the reasons above, we propose a robust closed-loop cross-layer
design
for OFDM systems where no explicit CSIT knowledge is needed at the base
station. The cross-layer power allocation, user assignment as well
as rate allocation are adaptive to the built-in 1-bit ACK/NAK feedbacks
\cite{Bhargava98a} \cite{Bhargava98b} \cite{Bhargava01} from the
selected users. Being built in at the link layer of most wireless
systems and hence, the ACK/NAK feedbacks add no incremental cost
to the proposed closed-loop design. Moreover, since the cross-layer solution is
driven by the ACK/NAK feedbacks, it introduces robustness on the cross-layer
performance with respect to uncertainty at the CSIT and propagation parameters.
These robustness cannot be obtained by utilizing explicit limited CSIT
feedback. However, there are several challenges in solving the problem:

\subsection{Technical Challenges}
\subsubsection{Issues of packet errors}
Conventional cross-layer optimization
only consider sum ergodic capacity as the optimization objection. Ergodic
capacity only considers the b/s/Hz transmitted by the BS regardless of packet
errors.  As a result, ergodic capacity is a reasonable performance metric only
when the packet error is negligible (which is the case with perfect CSIT and
very
strong coding). However, in our case without CSIT, there is always systematic
packet errors (due to channel outage) and this cannot be alleviated by just
using strong coding. To accommodate packet errors, we have to use system goodput
(b/s/Hz successfully received by the mobiles) as our performance metric. Note
that goodput reduces to ergodic capacity in the case of no errors but in
general, to deal with goodput, we need to deal with the cdf of mutual
information (rather than the first order moment only) and this impose some
technique challenges to the problem. 

\subsubsection{Issue of the MDP complexity}
While the problem belongs to MDP,
it is well-known that there is usually no simple solution (even numerically)
using standard value-iteration and policy-iteration solutions (see details in
section \ref{section:MDP}). For instance, the MDP belongs to the class of
infinite state space and brute-force approach has exponential complexity in the
number of time slots $M$ and hence, they could not give useful  solutions.
Instead of brute-force solution, we exploit some special structure of the OFDM
and obtained a low complexity closed-form solution, which is asymptotically
optimal for sufficiently small PER target.

\subsubsection{Asymptotic Performance}
As pointed out, all existing solutions
are heuristic in nature and studied performance purely by simulations. This is
because of the challenging nature of the problem. In this paper, we shall derive
some asymptotic properties on the system performance so as to obtain some design
insights.

\subsection{Summary of Contributions}
We consider the downlink of a wireless
system with
a base station and $K$ mobile users over frequency selective fading
channels (OFDM). The base station shall adapt the downlink
rate, power and user selection in an OFDM system based on
the ACK/NAK feedbacks from the mobiles. To take into account of
potential packet errors due to channel outage, we consider an average
system goodput which measures the number of bits successfully
transmitted as our performance measure. The robust cross-layer
design is modelled as a Markov Decision Process (MDP)
\cite{Doob53} \cite{Sarkar2004} \cite{Bettesh2006} \cite{Bhardwaj2007} with
power, rate and user selection policies
as the optimization variables so as to optimize the average system
goodput while maintaining a target PER.
It is well-known that MDP-based problems
\cite{Vijay05a}\cite{Vijay05b} always require complex value iteration
algorithms. However, in this paper, we shall derive a
simple closed-form solution for the MDP cross-layer problem which
is asymptotically optimal for sufficiently small target PER. The proposed
solution has low complexity and is suitable for
realtime implementation. It is also shown to achieve
significant performance gain compared with systems that do not
utilize the ACK/NAK feedbacks for cross-layer designs or
cross-layer systems that utilize very unreliable CSIT for
adaptation with mismatch in CSIT error statistics. Furthermore, since the
ACK/NAK feedbacks are generated
by the mobiles based on CRC checking after packet detection, the
proposed closed-loop cross-layer scheme is very flexible in the
sense that it can automatically accommodate mobiles with different
receive sensitivities  in the RF or variations in the baseband
estimation and decoding algorithms. Hence, the proposed scheme
achieve significant goodput gain with built-in robustness against
channel fluctuations as well as variations across the capabilities
of different mobile receivers.
\section{A Review on Markov Decision Process}\label{section:MDP}
MDP has found applications in ecology, economics and communications engineering
since 1950 \cite{Puterman1994}.
MDP is a modeling tool which describes a sequential decision making process. It
is used to make the \emph{optimal} sequence of decisions where 
outcomes of the problem are partly random and partly depend on such
decisions. The advantage of
MDP is that it provides a systematic framework for analysis of
optimality, existence, dynamics and convergence of solutions.

A complete description of a MDP problem involves a {\em decision epoch}, a {\em
state space}, a {\em control policy}, a {\em state transition kernel} as well as
a {\em reward function}. The time line is first divided into decision epochs in
which the controller makes decisions on control actions  and the system receives
{\em rewards} at the decision epochs. Specifically, at the $m$-th decision
epoch,
the system occupies a state $s_m \in \mathbb{S}$ where $\mathbb{S}$ denotes the
state space. Based on the observation on the causal state sequence
$s_1,...,s_m$, the controller takes a control action $\alpha_m \in \mathbb{A}$
where $\mathbb{A}$ is the set of actions. A {\em control policy} $\pi$ is
defined to be
the {\em set of actions} for all possible state sequence. Based on the action
$\alpha_m$ and the current state $s_m$, the system receives a reward $R(s_m,
\alpha_m)$
and moves to the next state $s_{m+1}$ according to the {\em state transition
probability kernel} $P(s_m, \alpha_m,s_{m+1})$. The optimization problem is to
find
the optimal control policy so as to maximize the total rewards: $\arg\max_{\pi}
\sum_{m=1}^M R(s_m, \alpha_m)$. As a result, a MDP problem can be
characterized by  the tuple $(\mathbb{S}, \mathbb{A}, P(.,.,.)), R(.,.))$. One
reason why the MDP problem is difficult is due to the huge dimensions of the
variable, namely the entire policy space $\pi$. As a result, a key step in
solving the MDP is known as {\em divide-and-conquer}. Specifically, instead of
optimizing for the entire problem, it can shown that the MDP can be solved by
optimization of actions $\alpha_m$ on a per-stage basis.  

There has been a lot of in-depth analysis of MDP \cite{Puterman1994}
\cite{Altman1998} and different branches of the problem. Different analysis are
needed for finite state space problems v.s. infinite state space problems;
finite horizon problems v.s. infinite horizon problems; {\em unconstrained} MDP
v.s. \emph{constrained MDP} etc. By constrained MDP, we mean that the problem
has one or more constraints on the feasible policy space $\pi$. Constrained MDP
problems are closely related to communication problems \cite{Altman1998} such as
power and rate control problems with an average delay
constraint \cite{Djonin2007}; scheduling problems involving routing in ad-hoc
networks \cite{Perkins1999} or handoff problems \cite{Wong2001}. For example, in
\cite{Wong2001}, the authors optimized the occurrence of path optimizations for
inter-switch handoffs in wireless ATM networks. The expected total cost per
call, including the switching/ handoff cost and signaling costs, is modeled as a
infinite-horizon semi-Markov decision process \cite{Rasanen2006} with discount
rate. This expected total cost is the objective function to be minimized. At
each decision epoch, the decision maker can choose to do path optimization or
not which is modeled in the action set. Using divide-and-conquer principle, the
MDP problem can be solved
using \emph{value iteration algorithm} or policy iteration
algorithm \cite{Feinberg2002}. The model is then
extended to have QoS constraints.

This paper is outlined as follows. The channel model is firstly
presented in section \ref{section:channelmodel}. In
section \ref{section:problemformulation}, the problem formulation is given
as a cross-layer optimization problem and a MDP problem. The conventional
 solutions of MDP is provided at the end of section
\ref{section:problemformulation}. The proposed solution, which is
asymptotically optimal, is presented in
section \ref{section:proposedsolution}. Simulation results are analyzed in the
section \ref{section:simulation}. Section \ref{section:conclusion} presents the
conclusion.
\section{Channel Model}\label{section:channelmodel}
\begin{figure}
\begin{center}
	\includegraphics[width=9cm]{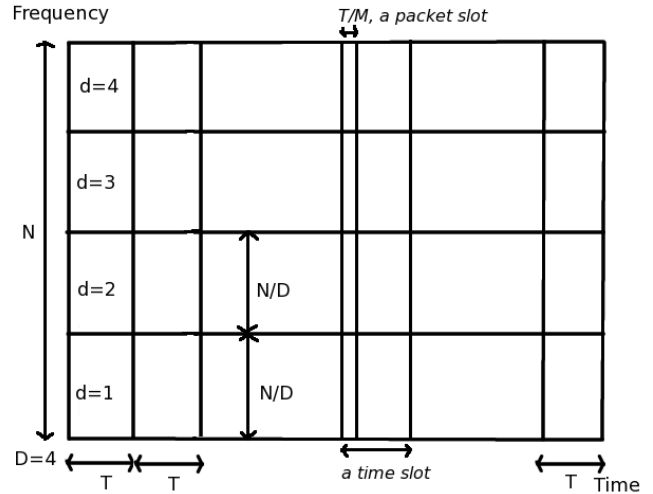}
	\caption{The channel model is represented graphically. In the
frequency domain, assume $D=4$ frequency blocks within $N$ subcarriers, there
are $\frac{N}{D}$ subcarriers in each frequency block and have the same
frequency gains. In the time domain, channel remains unchanged within $T$
seconds: a time slot. $M$ packets are transmitted in a time slot. Each packet
consume $\frac{T}{M}$ seconds: a packet slot.}\label{figure:slot_diagram}
\end{center}
\end{figure}

We consider a downlink cross-layer scheduling problem in a
frequency selective, block fading (in frequency) and quasi static (in
time) channel. The bandwidth is divided into $D$ frequency blocks.  The fading
gain in each frequency block is flat. With the
use of OFDM, the fading of each frequency block is independent to
other frequency blocks. Also, in the time domain, we assume that
the channel remains quasi-static for a period of time $T$ seconds and we call
this
a time slot. Thus, the fading gains on each frequency block remain the same
throughout a \emph{time slot}. Within a time slot, we send $M$
packets which occupy the same amount of time, a \emph{packet slot},
$\frac{T}{M}$ seconds. From now on, the names
packet slot and slot are used interchangeably. With frequency
block fading, there are $N$ frequency sub-carriers in which
$\lfloor \frac{N}{D} \rfloor$ frequency sub-carriers having the
same fading gains form a block and there are $D$ blocks in total.
The fading gain represented by each frequency block is assumed to be
independent of the other blocks. The model is summarized in figure
\ref{figure:slot_diagram}. 

Denote the number of users in the systems by $K$. Each user $k$ sees a vector
channel $\bar{h}_k=[h_{k,1}, \ldots, h_{k,D}]$ where $h_{k,j}$ is the channel
power of frequency block $j$ of user $k$. Stacking all vector channels, we have
a channel power matrix $H$.
\begin{equation}
H=\left(
\begin{array}{c}
\bar{h}_1 \\
\bar{h}_2 \\
\vdots \\
\bar{h}_K
\end{array}\right)
=\left(%
\begin{array}{ccccc}
  h_{1,1} & h_{1,2} & \cdots & \cdots & h_{1,D} \\
  h_{2,1} & h_{2,2} & \cdots & \cdots & h_{2,D} \\
  \vdots & \ddots &  &  & \vdots \\
  h_{K,1} & h_{K,2} &  & \ddots & h_{K,D} \\
\end{array}%
\right)
\end{equation}
Note that each entry $h_{k,d}$
is exponentially distributed with unit mean and variance. Denote
the ACK/NAK feedback from each user $k$ during packet slot
$m$ by $v_{k,m}$. Then,
\begin{equation}\label{eqt:ACK_NAK1}
v_{k,m}=\left\{%
\begin{array}{ll}
    1, & \hbox{ACK is received from user $k$ in slot $m$;} \\
    0, & \hbox{NAK is received from user $k$ in slot $m$.} \\
\end{array}%
\right.
\end{equation}
where ACK is received when the packet $m$ is successfully decoded
and NAK is received when the packet $m$ has error.

\begin{figure}[!h]
\begin{center}
  \includegraphics[width=9cm]{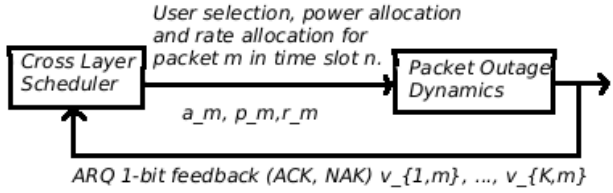}\\
  \caption{Closed Loop Cross-Layer Scheduler.
The user, power and rate optimization at the BSs is
solely based on the 1-bit feedbacksfrom MSs.
}\label{figure:scheduler_structure}
\end{center}
\end{figure}

The closed-loop cross-layer scheduler is as shown in figure
\ref{figure:scheduler_structure}. There are three optimization
parameters, namely the user selection $a_m$ , power level
$p_m$ and rate $r_m$ . The parameters are determined for each packet $m$. At the
receiver side, \emph{each user} $k$
would decode the packet and send a 1-bit ACK/NAK feedback $v_{k,m}$ to the
transmitter. In $m$-th packet slot, the maximum achievable rate in bits is 
\begin{equation}\label{eqt:capacity}
c(p_m,\bar{h}_{a_m})=\frac{NT}{DM} \sum_{d=1}^{D} \log_2 (1+
\frac{p_m h_{a_m,d}}{N})
\end{equation}
where noise power is normalized to be one.

Now, we can rewrite equation (\ref{eqt:ACK_NAK1}) mathematically,
\begin{equation}\label{eqt:ACK_NAK2}
v_{k,m}=\left\{%
\begin{array}{ll}
    1, & r_m \leq c(p_m,\bar{h}_{a_m}); \\
    0, & r_m \geq c(p_m,\bar{h}_{a_m}). \\
\end{array}%
\right.
\end{equation}

\begin{figure}
\begin{center}
		\includegraphics[width=9cm]{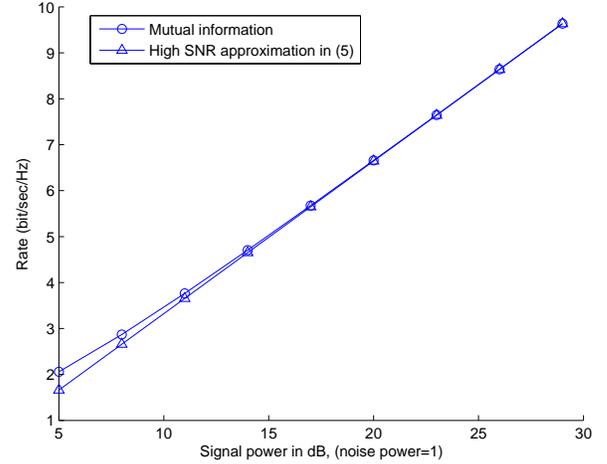}\\
  \caption{Rate difference between mutual
information and its approximation in
\eqref{eqt:cap_high_snr}. The difference is
less than 2\% in common operating region, between 10 to 30
dB.}\label{figure:mut_info_app_diff}
\end{center}
\end{figure}

In high SNR environment, the maximum bits per packet slot in
equation (\ref{eqt:capacity}) can be approximated by
\begin{eqnarray}\label{eqt:cap_high_snr}  
\nonumber & & c(p_m,\bar{h}_{a_m})= \frac{NT}{DM} \sum_{d=1}^{D}\log_2
(1+\frac{p_m h_{a_m,d}}{N})\\
 \nonumber & & \underset{high SNR}{\longrightarrow}
\frac{NT}{DM} \left(\sum_{d=1}^{D} \left( \log_2 \left(\frac{p_m}{N} \right)
\right) + \log_2( X_{a_m})\right)\\
 & & =c(p_m,X_{a_m}) 
\end{eqnarray}
where $X_{k}=\prod_{d=1}^D h_{k,d}$. This approximation 
significantly reduces the complexity of the system as the D-dimensional channel
power gain vector is replaced by a scaler. In figure
\ref{figure:mut_info_app_diff}, we show the difference between the maximum bits
per packet slot and its approximation in \eqref{eqt:cap_high_snr}. The
approximation error is less than 2\% when the SNR is around 10dB. 

Define the cumulative density function (CDF) of the random
variable $X_{k}$ to be
\begin{equation}\label{eqt:phi}
\phi(\chi)=Pr(X_k \leq \chi)
\end{equation}
which can be computed offline. Note that $X_k$ is unknown to the transmitter
which updates the set of all possible values of $X_k$ in each packet slot $m$ by
the feedback $v_{k,m}$. The set of all possible values $X_{k}$, based on
information received through feedbacks before packet slot $m$, is 
\begin{equation}\label{eqt:channel_power_set}
\mathbb{X}_{k,m+1}=\left\{%
\begin{array}{ll}
    \mathbb{X}_{k,m}\bigcap \left\{ X_k: c(p_m,X_k) \geq r_m \right\}, &
v_{k,m}=1; \\
    \mathbb{X}_{k,m}\bigcap \left\{ X_k: c(p_m,X_k) <  r_m \right\}, &
v_{k,m}=0. \\
\end{array}%
\right.
\end{equation}
For example, at packet slot 1, $m=1$, the set of real channel power gains
for $\mathbb{X}_{k,1}$ is all real numbers $\mathbb{R}^+$. A pair of power
and rate $(p_1,r_1)$ is selected. A packet is broadcasted with power $p_1$ and
rate $r_1$. At the end of packet slot 1, ACK/NAK feedbacks $v_{k,1}$ for all
users $k$ are
received. $\mathbb{X}_{k,2}, \forall k$ are then updated using 
\eqref{eqt:channel_power_set}. At the end of packet slot 2, $\mathbb{X}_{k,3}$
are
updated accordingly and so on.
Note that the set $\mathbb{X}_{k,m}$, as described in 
\eqref{eqt:channel_power_set}, would solely depend on the causal
power allocation, rate allocation and ACK/NAK feedbacks from the
users.
\section{Problem Formulation}\label{section:problemformulation}
This section is targeted to reveal the mathematical description of
the optimization problem. The problem is best explained by first
writing down the optimization variables which are the power, rate
and user selection policies defined in the following. We
would then provide the mathematical expression of the system goodput
which is the optimization objective in this paper. A problem
statement and its corresponding mathematical representation are
provided. A subsection is given here to explain the transformation
of the optimization problem to a MDP problem.
\subsection{Problem formulation as a cross-layer optimization problem}
For simplicity, denote the causal user assignments, rate sequence and power
 sequence from slots 1 to
$m-1$ by $A_m=(a_1,a_2,\ldots,a_{m-1}),$
$R_m=(r_1,r_2,\ldots,r_{m-1})$ and
$P_m=(p_1,p_2,\ldots,p_{m-1})$ respectively. Also, denote the causal ACK/NAK
feedbacks
for slots 1 to $m-1$ from all users by the matrix $V_m$
\begin{eqnarray}\label{eqt:causal_feedbacks}
\nonumber V_m &=& \left(%
\begin{array}{ccccc}
  v_{1,1} & v_{1,2} & \cdots & \cdots & v_{1,m-1} \\
  \vdots & \ddots &  &  & \vdots \\
  v_{K,1} & v_{K,2} &  & \ddots & v_{K,m-1} \\
\end{array}%
\right)\\
&=&\left(%
\begin{array}{c}
  \bar{v}_{1}^m  \\
  \vdots \\
  \bar{v}_{K}^m\\
\end{array}%
\right)
= \left( \bar{v}^1, \bar{v}^2, \ldots, \bar{v}^{m-1} \right)
\end{eqnarray}
\begin{Def}[Power Allocation Policy]
A power allocation policy
\begin{equation}
\mathcal{P}=\left\{\left(p_{m} \right)_{V_m}: \sum_{m=1}^M
p_{m} = P_0\right\}
\end{equation}
is defined as the set of all power allocation at the $m$-th packet
slot where $m\in [1,M]$. The subscript notation $(.)_{V_m}$
denotes that the power allocation at the $m$-th packet slot is a
function of the ACK/NAK feedbacks  up to the $(m-1)$-th packet slot $V_m$. The
power allocation policy $\mathcal{P}$ is restricted by the total
power constraint $P_0$. 
\end{Def}
Similarly, we define the rate allocation policy and user selection
policy.
\begin{Def}[Rate Allocation Policy]
A rate allocation policy
\begin{equation}
\mathcal{R}=\left\{ (r_m)_{V_m}: r_m \in
\mathbb{R}^+\right\}
\end{equation}
is defined as the set of all rate allocation at $m$-th packet slot
where $m \in [1,M]$ and $\mathbb{R}^+$ is the set of all positive real numbers.
The policy is determined by causal
ACK/NAK feedbacks up
to slots $m-1$. 
\end{Def}

\begin{Def}[User Selection Policy]
A user selection policy
\begin{equation}
\mathcal{A}=\left\{ (a_m)_{V_m}: a_m \in
\{1,\ldots,K\}\right\}
\end{equation}
is defined as the set of all user selection at $m$-th packet slot
where $m \in [1,M]$. The policy is determined by the causal ACK/NAK feedback
sequences up
to slots $m-1$. The user selection at $m$-th packet slot $a_m$ denotes the index
of user selected.
\end{Def}

Let the feedback of user $a_m$ at packet slot $m$ in time slot $z$ be
$v_{a_m,m}(z)$. The number of packet errors in time slot $z$ equals to the
sum of packet errors of the $M$ packets sent within time slot $z$:
$\sum_{m=1}^M (1- v_{a_m,m}(z))$. 
The total number of packet errors in $Z$ time slots is
$\sum_{z=1}^Z \sum_{m=1}^M (1-v_{a_m,m}(z))$. Thus, the packet error rate
averaged over time slots is
\begin{equation}
	P_e = \lim_{Z \rightarrow \infty} \frac{1}{MZ} \sum_{z=1}^Z
\sum_{m=1}^M (1-v_{a_m,m}(z)).
\end{equation} As the channel gain remains quasi-static within a time slot and
is
independent of that in other time slot, the averaged packet error rate can be
written as the expectation of number of packet errors within a time slot over
channel realizations.(We drop the notation of time slot $z$)
\begin{equation}
	P_e = \mathbf{E}_{H} \frac{1}{M} \sum_{m=1}^M (1-v_{a_m,m})
\end{equation} where $\mathbf{E}_{H}(.)$ denotes expectation over the random
variable $H$.

Note that the packet error rate can be simplified as follows.
\begin{equation}
P_e = Pr(c(p_m,X_{a_m})< r_m)
\end{equation}
The average system goodput
$\bar{G}$ (averaged over ergodic samples of time slots) is given by:
\begin{eqnarray}\label{eqt:avg_system_goodput}
\nonumber \bar{G}\left(
\mathcal{P},\mathcal{R},\mathcal{A}\right)&=&\mathbf{E}_H
\left\{ \sum_{m=1}^M v_{a_m,m}r_m\right\}\\
&=& \sum_{m=1}^M Pr(c(p_m,X_{a_m})>r_m) r_m .
\end{eqnarray}

In most wireless systems, a target packet error
rate (PER) is assigned due to various application
requirements. Let $\epsilon$ be that PER. For example, the PER,
$\epsilon$, is of the order of $10^{-2}$ for voice
applications. The relation between
$\mathbb{X}_{k,m}$ and $\epsilon$ \eqref{eqt:cap_high_snr} is given by
\begin{eqnarray}\label{eqt:quality}
\nonumber 1-\epsilon &=& Pr(c(p_m,X_{a_m}) \geq r_m |
\mathbb{X}_{a_m,m})\\
&=&Pr(X_{a_m}\geq \theta_m |\mathbb{X}_{a_m,m} )
\end{eqnarray}
where 
\begin{equation}\label{eqt:theta}
\theta_m= \left( \frac{N}{p_m} \right)^D 2^{\frac{
r_m DM}{NT}}.
\end{equation}

To conclude, the cross-layer optimization problem can be
formulated as
\begin{Problem}[Cross-layer formulation]\label{Prob:cross_layer_G}
Determine the optimal power allocation policy $\mathcal{P}$, rate
allocation policy $\mathcal{R}$ and user assignment $\mathcal{A}$
so as to maximize the average system goodput
$\bar{G}(\mathcal{P},\mathcal{R},\mathcal{A})$ subject to the
target PER requirement $1- \epsilon=
Pr\left( X_{a_m}\geq \theta_m | \mathbb{X}_{a_m,m}\right)$ and the total power
constraint $\sum_{m=1}^M p_m \leq P_0$.
\end{Problem}

The optimization problem above is difficult to solve due to the
huge dimension of variables involved. Yet, we shall illustrate
below that the total system goodput $\bar{G}$ can be expressed
recursively and hence, the problem above can be expressed as a
Markov Decision Problem. Define $F_m(\bar{P}_m,W_{m-1})$ to be the maximized
goodput
sum from slot $m$ to $M$ (from packet slot $m$ to the last packet slot) subject
to power constraint $\bar{P}_m$ and \emph{causal} power
allocations, rate allocations and feedbacks from users i.e. 
\begin{equation}\label{eqt:opt_problem_a}
F_m(\bar{P}_m,W_{m-1})= \underset{\mathbf{p}_m,\mathbf{r}_m,\mathbf{a}_m}{max}
\mathbf{E}_{H} \left\{ \sum_{i=m}^M v_{a_i,i} r_i \right\}
\end{equation} where $W_{m-1}=\left( V_{m-1}, A_{m-1}, \Theta_{m-1}= (\theta_1,
\ldots,
\theta_{m-1})\right)$ and $\mathbf{p}_m$ denotes the vector of power allocation
from
$p_m$ to $p_M$. Similar notations apply to $\mathbf{r}_m$ and $\mathbf{a}_m$.
The maximization is subject to the PER
requirement $Pr(X_{a_m} \geq \theta_m | \mathbb{X}_{a_m,m})=1-\epsilon$ and the
total power constraint $\sum_{i=m}^M p_i \leq \bar{P}_m$. We first have the
following lemma about $F_m(\bar{P}_m,W_{m-1})$.

\begin{Lem}\label{Lem:F_recur}
 $F_m(\bar{P}_m,W_{m-1})$ can be espressed recursively as
\begin{eqnarray}\label{eqt:opt_problem}
 & & F_m(\bar{P}_m,W_{m-1})\\
&=&\nonumber \underset{p_m,r_m,a_m}{max}
\left\{ (1-\epsilon)r_m  +
\mathbf{E}_{V_m}\left[F_{m+1}(\bar{P}_m-p_m,W_m)\right]\right\}.
\end{eqnarray} 
\end{Lem}
\begin{proof}
See subsection \ref{Appendix:resur_goodput} in appendix.
\end{proof}
Note that the maximization variables are $p_m,r_m,a_m$, the
power, rate and user selection in packet slot $m$, instead of the selections
from slot $m$ till the last slot. As a result, this facilitate the
divide-and-conquer approach to the original optimization problem in
\eqref{Prob:cross_layer_G}. 

From (\ref{eqt:avg_system_goodput}), the maximized system goodput is 
\begin{equation}
\bar{G}^*(\mathcal{P},\mathcal{R},\mathcal{A})=
\underset{\mathbf{a}_m,\mathbf{p}_m,\mathbf{r}_m}{max} \sum_{m=1}^M
\mathbf{E}_H \left\{ v_{a_m, m}\right\} r_m.
\end{equation}
By definition of $F_m$ in equation (\ref{eqt:opt_problem_a}), the optimized
goodput is 
\begin{eqnarray}\label{eqt:barG}
\bar{G}^*(\mathcal{P},\mathcal{R},\mathcal{A})&=&F_1(P_0,W_0)\\
\nonumber \mbox{subject to} & & \sum_{m=1}^M p_m \leq P_0 \\
\nonumber & & Pr(X_{a_m} \geq \theta_m | X_{a_m} \in \mathbb{X}_{a_m,m})= 1-
\epsilon\\
\nonumber & & \theta_m= \left( \frac{N}{p_m}\right)^D 2^{\frac{r_m DM}{NT}}
\end{eqnarray}
where $W_0$ is a empty set. As a result, the optimized system goopdut
$\bar{G}^*(\mathcal{P},\mathcal{R},\mathcal{A})=F_1(P_0,W_0)$ can be
obtained
recursively from equation (\ref{eqt:opt_problem}). We shall eleborate the
recursive solution in the following sections.

\subsection{Problem Formulation as a Markov Decision Process}
As explained in Section II, a MDP problem is characterized by the tuple
$(\mathbb{T},\mathbb{S},\mathbb{A},P(s,\alpha, s'),R(s,\alpha))$. In our case,
the decision epochs of the base station $\mathbb{T}=\{1,2,...,M\}$ corresponds
to the scheduling slots. In the following, we shall discuss the association of
our cross-layer optimization problem with the MDP tuple, namely the state space
$\mathbb{S}$, action space $\mathbb{A}$, state transition kernel as well as
the per-stage reward function. Based on that, we shall formally recast the
problem into an MDP. 

\begin{itemize}

\item {\bf State Space Association} 
With $\Theta_m=[\theta_1, \ldots, \theta_m]$, define $U(\Theta_m, \bar{v}_k^m)$
and $ L(\Theta_m, \bar{v}_k^m)$ to be the
upper bound and lower bound of CSI which is some information gathered by the
ACK/NAK feedbacks $\bar{v}_k^m$ and $\theta_m$ in equation (\ref{eqt:theta}).
The state space, $\mathbb{S}$, is a collection of the following vectors $s$.
\begin{equation} 
s=(L(\Theta_m, \bar{v}_k^m), U(\Theta_m,\bar{v}_k^m), \theta_m,
\bar{P}_m, \bar{R}_m, \vec{s}^{(ACK)}, \vec{s}^{(NAK)})
\end{equation} where
$\bar{P}_m$ is the remaining power;  $\bar{R}_m$ is the sumrate from slot $m$ to
$M$, $\vec{s}^{(ACK)}$ and $\vec{s}^{(NAK)}$
are the \emph{pointers} to the states if ACK:$v_m=1$ and NAK: $v_m=0$
respectively.

The CSI can take all possible real
values and therefore make the state space $\mathbb{S}$ infinite. However, as
illustrated in an example in the following subsection, the decision tree built
by state transitions in our problem is a lot smaller in size.

\item {\bf Action Space and Policy Association} 
The action taken at each state $s$ consists of  the selection of power $p_m$,
transmission rate, $r_m$, and the user selection, $a_m$. The set of
possible actions $\mathbb{A}$ at every state $s$ is independent of decision
epoch m and it is given by:
\begin{eqnarray}
& &\mathbb{A}=\mathbb{A}_{s,m} = \left\{ (p_m, r_m, a_m) \in \right.\\
& & \nonumber  \left. \left\{ p \in
\mathbb{R}^+ : p \leq P_0\right\} \times \mathbb{R}^+ \times \left\{ 1, \ldots,
K\right\} \right\}.
\end{eqnarray}
\item {\bf State Transition Kernel Association} The transition probability
$P(s,\alpha , s')$ is a real value function which
maps $\left\{ \mathbb{S} \times \mathbb{A} \times \mathbb{S} \right\}$ to
$[0,1]$. In our case, the probability of going from state $s$
to state $s'$ by action $\alpha \in \mathbb{A}$ is time invariant. 

In each decision epoch, $m$, a selection of actions, $\alpha_m$, takes place,
meaning that the base station selects the power $p_m$ and the transmission rate
$r_m$ to user $a_m$. After every user $k$ receives the packet, each of them
would decode the packet header and transmit a 1-bit feedback to base station,
$v_{k,m}$. This 1-bit feedback carries the information of ACK (1) or NAK (0).
The transition probability captures the probability of such ACK (1) or NAK (0)
and would take the system to a different state. For instance, the current state
is denoted by $s$; the state after receiving ACK $s^a$; the state after
receiving NAK
$s^n$. The probability of receiving ACK is $\mathcal{P}_a$ and that of
NAK is
$1-\mathcal{P}_a$. The action taken is $\alpha$. We have
\begin{eqnarray}
P(s,\alpha, s^a)=\mathcal{P}_a;\\
P(s,\alpha, s^n)=1-\mathcal{P}_a.
\end{eqnarray}
And
\begin{equation}
\sum_{s' \in \mathbb{S}} P(s, \alpha, s')=1
\end{equation}
The state transition probability is described in equation
\eqref{eqt:trans_prob}
\begin{figure*}[!t]
\normalsize
\setcounter{MYtempeqncnt}{\value{equation}}
\setcounter{equation}{30}
\begin{equation}\label{eqt:trans_prob}
P(s, \alpha, s') = P(\theta_{m+1}=\mathit{\theta}' | \theta_m=
\mathit{\theta}, \alpha_m= \alpha)= \begin{cases}
      \epsilon,&  Pr(X_k > \mathit{\theta}' | X_k > L(\Theta_m, \bar{v}_k^m),
X_k<U(\Theta_m,\bar{v}_k^m))= \epsilon \\
      1- \epsilon, & Pr(X_k > \mathit{\theta}' | X_k > L(\Theta_m, \bar{v}_k^m),
X_k<U(\Theta_m, \bar{v}_k^m))=1- \epsilon\\
      0 & \text{otherwise.} 
 \end{cases}
\end{equation}
\begin{equation}\label{eqt:trans_prob2}
P(s, \alpha, s')=\begin{cases}
      \epsilon,&  \phi (\theta_{m+1}=\mathit{\theta}')= (1- \epsilon) \phi
(U(\Theta_m, \bar{v}_k^m))+
\epsilon\phi (L(\Theta_m, \bar{v}_k^m)) \\
      1- \epsilon, & \phi (\theta_{m+1}=\mathit{\theta}')=  \epsilon \phi
(U(\Theta_m, \bar{v}_k^m))+
(1-\epsilon) \phi (L(\Theta_m, \bar{v}_k^m)) \\
      0 & \text{otherwise.} 
 \end{cases}
 \end{equation}
\begin{equation} \label{eqt:MDP_constraints}
   P(s_m, \alpha, s_{m+1})=\begin{cases}
      \epsilon,&  \phi (\theta_{m+1})= (1- \epsilon) \phi (U(\Theta_m,
\bar{v}_k^m))+ \epsilon\phi (L(\Theta_m, \bar{v}_k^m)) \\
      1- \epsilon, & \phi (\theta_{m+1})=  \epsilon \phi (U(\Theta_m,
\bar{v}_k^m))+
(1-\epsilon) \phi (L(\Theta_m, \bar{v}_k^m)) \\
      0 & \text{otherwise.} 
 \end{cases}
 \end{equation}
\setcounter{equation}{\value{MYtempeqncnt}}
\hrulefill
\vspace*{4pt}

\end{figure*}

in which $\mathit{\theta}'$ is the third element in $s'$ and
$\mathit{\theta}$ is the third element in $s$.
The upper and lower bound of CSI would be modified according to
the ACK/NAK
feedbacks received. After updating the bounds, the probability of ACK, which
is equal to the probability of the event that the channel power $X_k$ lies
between the lower bound and state $\mathit{\theta}'$, has to equal $1-
\epsilon$, as dictated
by the error constraint. Evaluate the probability, we have equation
\eqref{eqt:trans_prob2}.

\item {\bf Per-stage Reward} To
decide which actions in $\mathbb{A}$ should be carried out, we would need a
decision rule $d_m$. The decision rule $d_m$ is a history-dependent function.
Define the history $\delta_m$ to be a vector of past states, actions and
feedbacks.
\begin{equation}
\delta_m=[s_1, \alpha_1,  \ldots, s_{m-1}, \alpha_{m-1}, s_m ]
\end{equation}
The recursive relation is therefore
\begin{equation}
\delta_m = [ \delta_{m-1}, \alpha_{m-1},s_m].
\end{equation}
Denote the set of all histories by $\Delta_m$. Note that
\begin{eqnarray}
\Delta_1&=& \mathbb{S}\\
\nonumber \Delta_2 &=& \mathbb{S} \times \mathbb{A} \times  \mathbb{S}\\
\nonumber &\vdots&\\ 
\nonumber \Delta_m &=& \mathbb{S} \times \mathbb{A} \times  \cdots \times
\mathbb{S}\\
\nonumber &=& \Delta_{m-1} \times \mathbb{A} \times  \mathbb{S}
\end{eqnarray}
The history dependent rule $d_m$ maps $\Delta_m$ to $\mathbb{A}$.

A control policy is a plan specified by a sequence of decision rules. A
control policy $\pi$ is 
\begin{equation}
\pi= (d_1, d_2, \ldots, d_M), \; \; d_i \in \Delta_i, i=1, \ldots, M
\end{equation}
\addtocounter{equation}{3}
The per-stage reward function is 
\begin{equation}
	R(s_m, \alpha_m)=  \left\{ \begin{array}{cc}
P(s_m, \alpha, s_{m+1,a}) r_m & if \; v_m=1;\\
0 & if \; v_m=0;
\end{array} \right.
\end{equation}
where $s_{m+1,a}$ denotes the state at slot $m+1$ if $s_m$ is reached at slot
$m$ and action $\alpha_m$ is taken.

\end{itemize}
\begin{Problem}[The MDP formulation]\label{problemformulation:mdp}
The MDP problem is defined as a maximization problem of the reward function, in
our case, the system goodput
$F_1(P_0, W_0)$. Thus, the problem statement is, with slightly abuse of notation
\begin{equation}
\underset{\pi}{max} \left\{ \sum_{m=1}^M
R(s_m,\alpha_m) \right\}
\end{equation}
such that $\forall m=1,\ldots,M, s_m, s_{m+1} \in \mathbb{S}$,
$\alpha_m \in \mathbb{A}, r_m \in \mathbb{R}^+$ and equation
\eqref{eqt:MDP_constraints} is satisfied.
\end{Problem}

\begin{figure}[!ht]
\begin{center}
  \includegraphics[width=9cm]{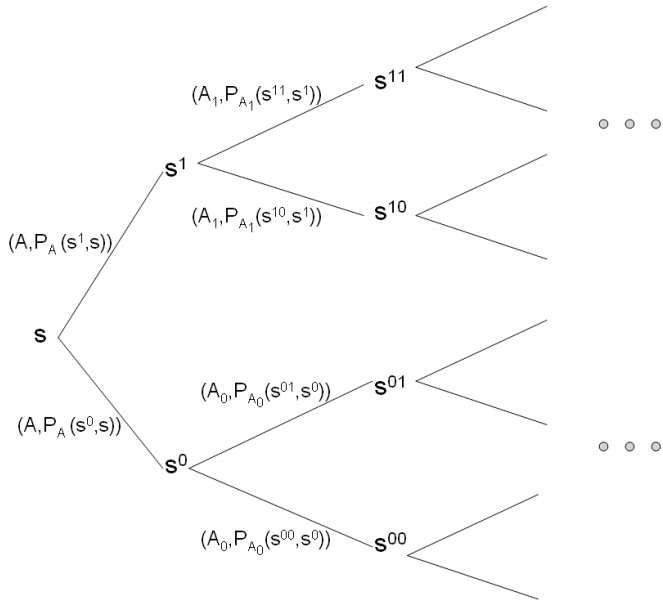}
  \caption{A state transition diagram example.
With only 2 possible outcomes at each state (node),
the state space (the number of nodes) increases
exponentially, hence the problem size.}\label{fig:state_transition}
\end{center}
\end{figure}

\subsection{A State Transition Example}
To illustrate the state transition of a MDP, a state transition
diagram assigned with an initial state is given in figure
\ref{fig:state_transition} by only
drawing transition branches corresponding to the tuples of
scheduled action and the corresponding non-zero transition
probability. Note that this diagram only shows a fragment of the whole decision
tree because there are more than one possible initial state.

The decision tree has $O(|\theta_m|)^3 \times 2^M$ elements, where $|\theta_m|$
is the number of values $\theta_m$ can take. In other words, 
\begin{equation}
|\mathbb{S}|=O(|\theta_m|)^3).	
\end{equation}
 There are $|\theta_m|$ possible values of the lower bound $L(\Theta_m,
\bar{v}_k^m)$. For example, $L(\Theta_m, \bar{v}_k^m) \in \left\{y_1, \ldots,
y_{|\theta_m|} \right\}$ where $y_b < y_{b+1}$. For each value of lower bound
$y_b$, there are $|\theta_m|-b-1$ values of $U(\Theta_m, \bar{v}_k^m)$ and
$\theta_m$. Thus, the total number of possible states is $1^2+2^2+ \ldots +
|\theta_m|^2= O(|\theta_m|^3)$. 

With either positive or negative feedbacks, each
state can only branch to 2 possible next states. Assume that we start on one
of these states. The number of possible \emph{descendents}  would be equal to
the sum of the series $1+2+2^2+2^3+ \ldots + 2^{M-1}$ which is $2^M$. Thus, the
total number of nodes in the tree is $O(|\theta_m|)^3 \times 2^M$.

Denote the
elements in the state space $\mathbb{S}$ by
\begin{equation}
\mathbb{S}=\left\{s, \{s^0,s^1 \}, \{s^{00},s^{01},s^{10},s^{11}
\},\ldots, \{ s^{q_{M-1}}\}\right\}
\end{equation}
where $q_{M-1}$ denotes any possible binary sequence of length
$M-1$. The binary sequence represents the causal ACK or NAK
feedbacks received. For example, state $s^{00}$ represents that 2
NAKs have been received and state $s^{101}$ represents that the
first and the third transmission are correct and the second
transmission or guess is incorrect. The state $s^{q_i}$ is at the
$i$-th level of the tree  which means the
$(i+1)$-th packet
transmission (with the root being the zeroth level). In the
diagram, only transitions with
non-zero probability are drawn. The transition probability
corresponding to action $A_{q_i} \in \mathbb{A}$ from state
$s^{q_i}$ to state $s^{[q_i,0]}$, meaning that a NAK is received
at $(i+1)$-th packet transmission, is denoted by the probability
$P\left( s^{q_i}, A_{q_i},s^{[q_i,0]}\right)$. At each state
$s^{q_i}$, there are two possible transition branches

\begin{eqnarray}
& & ACK :\\
& &\nonumber \left( A_{q_i},P\left( s^{q_i},A_{q_i},s^{[q_i,1]}\right)\right)
=\left((p_{i+1},r_{i+1},a_{i+1}), 1-\epsilon\right)\\
& & NAK: \\
& & \nonumber \left( A_{q_i},P\left( s^{q_i},
A_{q_i},s^{[q_i,0]}\right) \right)=\left((p_{i+1},r_{i+1},a_{i+1}),
\epsilon\right).
\end{eqnarray}

\subsection{Conventional Solutions of MDP}
A conventional solution to a MDP consists of backward and forward recursions.
The backward recursions set up a huge searching tree/ table which would involves
dynamic programming. In the forward recursions, the system states evolve through
the tree. Here we adopted the Finite Horizon-Policy Evaluation Algorithm in
\cite{Puterman1994} for the backward recursion.

\begin{algorithm}
\caption{Conventional Finite Horizon- Policy Evaluation Algorithm}
\label{alg:DP}
\begin{algorithmic}[1]
\State Each node in the tree consists of following fields:
$(L,U,\theta_m,\bar{P}_m,\bar{R}_m,\vec{s}^{(ACK)},\vec{s}^{(NAK)})$.
\State Initialization: $m \leftarrow M$, $\forall L,U, p_M, \theta_M$ 
\Statex $F_M^* (p_M, \delta_M)= \underset{d_M(\delta_M) }{max} Pr(
c(p_M,\theta_M)>r_M)r_M$
\State if $m=1$, stop. Otherwise, go to step 4.
\State $m \leftarrow m-1$, $\forall s_m,p_m,\bar{P}_m,L,U$ 
\Statex Evaluate
$F_m^*(\bar{P}_m,\delta_m)= \underset{d_m(\delta_m)}{max} \left\{ P(s_m,\alpha,
s_{m+1})r_m \right.$ 
$+ P(s_m,\alpha, s_{m+1})F_{m+1}^* (\bar{P}_m-p_m, \delta_m |
v_{a_m, m}=1)$
$ \left. + (1-P(s_m,\alpha,s_{m+1})) F_{m+1}^*(\bar{P}_m-p_m,
\delta_m | v_{a_m, m}=0)\right\}$
such that the constraints in equation \eqref{eqt:MDP_constraints} are
satisfied and $ P(s_{m}, \alpha, s_{m+1})= 1- \epsilon$ 

\State $(p_m,a_m,r_m)$ are given by $d_m(\delta_m)$ obtained in step 4.
\State $\bar{R}_m=F_m^*(\bar{P}_m,\delta_m)$ which is the accumulated rate
of this node and its descendents.
\State $\vec{s}^{ACK},\vec{s}^{NAK}$ are computed in
\eqref{eqt:MDP_constraints} 
\end{algorithmic}
\end{algorithm}

 After building up a table in
backward recursion using algorithm \ref{alg:DP}, from $m=M \rightarrow 1$, we
established a large binary tree with each node represents a particular
estimate of channel power and each branch corresponds to an ACK/NAK
feedback. Each path from the root to the leaves corresponds to a sequence
of estimates and the corresponding feedbacks. In  \emph{Online Evolution}
(algorithm \ref{alg:DP_on}), we read this tree from the root and traverse down
to
the leaves. Each packet is transmitted with parameters marked in that node and
a new node is reached according to the ACK/NAK feedbacks.

\begin{algorithm}
\caption{Conventional Online State Evolution Algorithm}
\label{alg:DP_on}
\begin{algorithmic}[1]

\State Set $m=1$ and start state 
\Statex $s=(0, \infty,\theta_m,P_0,\bar{R}_m,\vec{s}^{(ACK)},\vec{s}^{(NAK)})$ 
\Statex where $\bar{R}_m$ is the maximum among the nodes with $L=0, U=\infty$.
\State If $m=M+1$, stop, otherwise go to step 3.
\State Transmit packets as prescribed by decision rule $d_m(\delta_m)$ computed
in algorithm \ref{alg:DP}.
\State Receive an ACK/NAK feedback $v_{k,m}$ from each user $k$.
\State Update the upper and lower bound of CSI.
\Statex $L = \theta_m \; \mbox{if } v_{a_m,m}=1$
\Statex $U = \theta_m \; \mbox{if } v_{a_m,m}=0$ 
\State Evolve to next state according to the bounds of CSI $\vec{s}^{(ACK)},
\vec{s}^{(NAK)}$ and feedbacks $v_{k,m} \forall k$.
\State $m+1 \leftarrow m$, go to step 2.
\end{algorithmic}
\end{algorithm}

Note that the drawback of such algorithm is that the requirement of memory is
huge as there are numerous possible states. In our problem, the state space is
infinite. Even if we discretize the state space as an approximation, the
complexity of the brute-force approach has exponential complexity in $M$ and
hence, could not give viable solutions. 
\section{Proposed Solutions}\label{section:proposedsolution}

The MDP can be 
solved by a backward recursion followed by a forward recursion. In
this section, we shall first elaborate the backward recursive
solution, namely the \emph{Optimal State Evolution}
followed by the forward recursion, namely the \emph{Online
Envolution}. Unlike conventional solution for MDP, we proposed a
simple closed-form solution which is
asymptotically optimal for sufficiently small PER. The proposed
solution only has complexity $O(M)$, which is in big contrast with brute-force
complexity $O(exp(M))$.

\subsection{Optimal State Evolution}
We illustrate how to combine the target PER $\epsilon$,
with the knowledge obtained from feedbacks to generate estimates of channel
power $\theta_m$.
Note that $\theta_m$ in equation (\ref{eqt:theta}) is always either $\sup
\mathbb{X}_{k,m}$ or $\inf \mathbb{X}_{k,m}$ as equation
(\ref{eqt:channel_power_set}) can
be rewritten as
\begin{equation}\label{eqt:channel_power_set_theta}
\mathbb{X}_{k,m+1}=\left\{%
\begin{array}{ll}
    \mathbb{X}_{k,m}\bigcap \left\{ X_k: X_k \geq \theta_m \right\}, &
v_{k,m}=1; \\
    \mathbb{X}_{k,m}\bigcap \left\{ X_k: X_k <  \theta_m \right\}, & v_{k,m}=0.
\\
\end{array}%
\right.
\end{equation}
The lower bound and upper bound of $\mathbb{X}_{k,m+1}$  
are
\begin{eqnarray}\label{eqt:bounds_update}
& & L(\Theta_m, \bar{v}_k^m)=\max \left\{ \theta_i : v_{k,i}=1, \;
1 \leq i \leq m \right\}\\
& & U(\Theta_m, \bar{v}_k^m)=\min \left\{ \theta_i : v_{k,i}=0, \;
1 \leq i \leq m \right\}.
\end{eqnarray}

Combine \eqref{eqt:quality} with the knowledge obtained from feedbacks:
\begin{equation}\label{eqt:quality_theta}
Pr\left( X_k \geq \theta_{m+1} | X_k \geq
L(\Theta_m,\bar{v}_k^m),X_k < U(\Theta_m,\bar{v}_k^m)\right)=1-
\epsilon
\end{equation}
Rearranging the terms in equation (\ref{eqt:quality_theta}), we
have the dynamics of $\theta_m$
\begin{Lem}
	At each packet slot $m$, the
estimate of channel power $X_{a_m}$ is
computed by the causal feedbacks
$\bar{v}_{a_m}^{m-1}$ and the lower
and uppwer bound of $X_{a_m}$ 
\begin{equation}\label{eqt:theta_dynamics}
\phi(\theta_m)=\epsilon \phi
(U(\Theta_{m-1},\bar{v}_{a_m}^{m-1}))+(1- \epsilon)\phi
(L(\Theta_{m-1},\bar{v}_{a_m}^{m-1}))
\end{equation} where $\phi(\theta_m)$ is the cdf of
$X_{a_m}$ \eqref{eqt:phi}.
\end{Lem}
\begin{proof}
	see section \ref{appendix:theta} in appendix.
\end{proof}

\subsection{User Selection}
Evaluate the expectation in $F_m(\bar{P}_m,W_{m-1})$ defined in
\eqref{eqt:opt_problem},
we obtain equation
(\ref{eqt:F_recursion}).
\begin{figure*}[!t]
\normalsize
\setcounter{MYtempeqncnt}{\value{equation}}
\setcounter{equation}{44}
\begin{equation}\label{eqt:F_recursion}
F_m(\bar{P}_m,W_{m-1})=\underset{p_m,r_m,a_m}{max}\left\{ (1-\epsilon)r_m +
(1-\epsilon) F_{m+1}(\bar{P}_m-p_m,W_m | v_{a_m,m}=1)+\epsilon
F_{m+1}(\bar{P}_m-p_m,W_m | v_{a_m,m}=0)\right\}
\end{equation}
\begin{equation}\label{eqt:F_recursion_short}
F_m(\bar{P}_m,W_{m-1})=\underset{p_m,r_m,a_m}{max}
\left\{ (1-\epsilon)r_m + (1-\epsilon)
F_{m+1}(\bar{P}_m-p_m,W_m|v_{a_m,m}=1)\right\}.
\end{equation}
\begin{equation} \label{eqt:decision_rule}
d_m(\delta_m)= \left(p_m=\frac{\epsilon \bar{P}_m}{1-
(1-\epsilon)^{M-m+1}},
r_m=\frac{NT}{DM}\log_2 \left( \left( \frac{p_m}{N}\right)^D \theta_m \right) ,
a_m= \underset{k}{\arg \max}L (\Theta_{m-1}, \bar{v}^{m-1}_k)\right)
\end{equation}
\setcounter{equation}{\value{MYtempeqncnt}}
\hrulefill
\vspace*{4pt}

\end{figure*}
Solving equation (\ref{eqt:F_recursion}), a stochastic programming
tree would be needed. Yet, as $\epsilon$ is small for practice,
the decision tree is reduced to equation \eqref{eqt:F_recursion_short}.

The complexity of the problem is reduced from exponential to linearity with $m$.
\addtocounter{equation}{3}
\begin{Lem}
The optimal user selection strategy 
\begin{equation}
a_m= \underset{k}{\arg \max} \; L(\Theta_{m-1}, \bar{v}_k^{m-1})
\end{equation} of \eqref{eqt:F_recursion_short} is

\end{Lem}
\begin{proof}
See subsection \ref{Appendix:opt_user} in appendix.
\end{proof}

\subsection{Power Allocation}
\begin{Lem}
The power allocation policy
\begin{equation}
p_m=\frac{\epsilon \bar{P}_m}{1- (1- \epsilon)^{M-m+1}}
\end{equation},
where $\bar{P}_m= P_0 -
\sum_{i=1}^{m-1} p_i$ is the remaining power at time $m$,  is an optimal policy
with respect
to optimization problem \eqref{eqt:F_recursion_short}.

\end{Lem}
\begin{proof}
See subsection \ref{Appendix:opt_power} in appendix.
\end{proof}

\subsection{Rate Allocation}
Given the causal feedback, power and rate information $W_m$ and
the channel estimate/state values $\theta_m$ in
(\ref{eqt:theta_dynamics_inverse}) at each slot $m$, the rate
allocation is computed by the following
\begin{equation} \label{eqt:rate}
r_m= \frac{NT}{DM} log_2 \left( \left( \frac{p_m}{N}\right)^D
\theta_{m}\right)
\end{equation}

\subsection{Online Evolution}
With new information, $v_{k,m-1}$ arrives in each slot $m$, we
proceed on the decision tree according to the updated upper and lower bounds of
CSI and the feedbacks. The set $\mathbb{X}_{k,m}$ is modified to
contain only the possible values of the channel power gain based on the
causal ACK/NAK feedbacks. $\mathbb{X}_{k,m}= \left\{ x:
L(\Theta_m,\bar{v}_k^m) < x< U(\Theta_m,\bar{v}_k^m)\right\}$ The transmission
parameters according to the decision rule are in equation
\eqref{eqt:decision_rule}.

\begin{figure}[!t]
  \begin{center}
  \includegraphics[width=9cm]{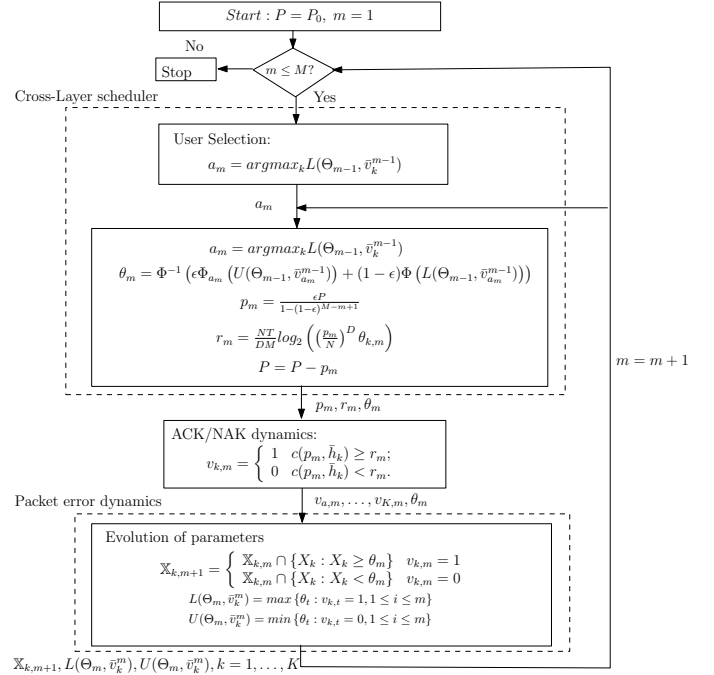}\\
  \caption{Structure and implementation of the proposed solution. }
  \label{figure:scheduler_math}
  \end{center}
\end{figure}

User $a_m$ is selected such that she
contains the largest possible channel power gain. As proved before, the
power allocation is static and solely depends on the total power
and the target error probability constraint. The data rate is adapted according
to channel estimate
$\theta_{m}$ and feedbacks $v_{a_m,m-1}$. The
online scheduling policy is illustrated in figure
\ref{figure:scheduler_math}.
\section{Asymptotic Analysis}
This section is devoted to prove that the goodput achieved in a
packet slot would be equal to the instantaneous mutual information of the slot
as if they were
perfect CSIT when the number of transmissions or number of packet
transmissions tends to infinity. In other words, there is zero
steady-state-error in the recursive solution.
To prove such claim, we would
need the following four theorems.

\begin{Lem}\label{lem:subset}
At packet slot $m$, the users selection set $\mathbb{K}_m$ denotes the set of
users who
have the largest potential channel power gains.
\begin{equation}\label{eqt:user_selection_set} 
\mathbb{K}_m=\left\{ k : L(\Theta_m,\bar{v}_k^m) > L(\Theta_m \bar{v}_{k'}^m) ,
\forall k' \not\in  \mathbb{K}_m \right\}
\end{equation}
The users selection set $\mathbb{K}_m$ at slot m is a subset of
$\mathbb{K}_{m-1}$.
\begin{equation}\label{eqt:user_selection_subset}
\mathbb{K}_m \subset \mathbb{K}_{m-1}
\end{equation}
The number of elements in $\mathbb{K}_m$ is
$|\mathbb{K}_m|$ which decreases with $m$.
\end{Lem}

\begin{proof}
See subsection \ref{Appendix:shrink_user_set} in appendix.
\end{proof}

\begin{Lem}\label{lem:real_in_bounds}
For all users $k$ in user selection set $\mathbb{K}_m$ at each slot $m$, the
channel power gains $X_k$ have lower bounds and upper bounds $L(\Theta_m,
\bar{v}_k^m)$ and $U(\Theta_m, \bar{v}_k^m)$.
\end{Lem}

\begin{proof}
See subsection \ref{Appendix:user_ch_bound} in appendix.
\end{proof}

\begin{Lem}\label{thm:gap_decrease}
Define the gap between the upper and lower bounds of channel power gains to be
$w_m= U(\Theta_m, \bar{v}_{a_m}^m)-L(\Theta_m, \bar{v}_{a_m}^m)$. $w_m$
monotonically decreases with $m$.
\end{Lem}

\begin{proof}
See subsection \ref{Appendix:bound_gap} in appendix.
\end{proof}

\begin{Lem}
When number of transmissions goes to infinity, the scheduled rate $r_m$ achieves
capacity of the system in perfect CSIT case. In the other words, the scheduled
rate $r_m$ is equal to the capacity achieved by selecting user which gives
highest capacity and using perfect CSIT.  Or mathematically, 
\begin{equation*} 
\underset{m \rightarrow \infty}{\lim} r_m= \underset{m \rightarrow \infty}{\lim}
\frac{NT}{DM} log_2 \left( \left( \frac{p_m}{N}\right)^D
\theta_{m}\right) = c(p_m, X_{a_m}).
\end{equation*}
\end{Lem}

\begin{proof}
See subsection \ref{Appendix:rate_cap} in appendix.
\end{proof}

\section{Results and Discussions}\label{section:simulation}
In this section, we would discuss the simulation results with the
following simulation settings. The bandwidth of the systems is 20
MHz which is divided into 64 subcarriers (N=64). Throughout these
subcarriers, there are $D$ group of independent subbands. The time slot $T =
0.1$ sec and we compared our proposed solution with two baselines. Specifically,
in baseline 1, we assume the BS has perfect CSIT and performs standard power
adaptation and hence, it serves as a goodput upper bound. In baseline 2, we
consider round robin scheduling which does not utilize any CSIT information and
hence, has very robust performance against CSIT errors. Note that the
performance of baseline 1 is obtained under perfect CSIT assumption and
therefore is not achievable. By comparing with baseline 1, we can guage how {\em
optimal} the proposed solution could achieve. Similarly, by comparing with
baseline 2 (which is a common approach in the absence of CSIT), we could guage
the potential performance advantage that can be captured by utilizing the
built-in ACK/NAK feedback flows.

\begin{figure}[!t]
\begin{center}
  \includegraphics[width=9cm]{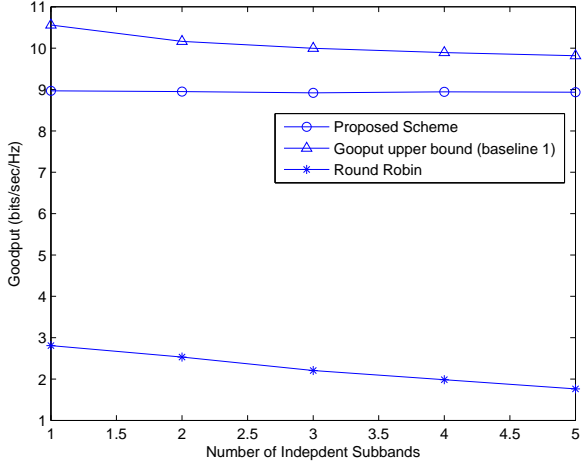}\\
  \caption{Average system goodput vs number of independent subbands with
transmit SNR=30dB, $P_0=24W, K=3, M=30, PER=0.05$.}\label{figure:numch}
\end{center}
\end{figure}

\subsection{Effects of Number of Independent Subbands}
In figure \ref{figure:numch}, the sum of goodput in 30 packets
transmitted is plotted against the number of independent subbands $D$ with
$P_0=24W, SNR=30 dB, K=3$ and target $PER = 0.05$. 
 
Note that our proposed solution achieved 85\% and 91\% of the performance upper
bound (baseline 1) when $D$ = 1 and 5 respectively. Compared with baseline 2
(RR), the proposed solution achieved very significant 500 \% goodput gain. This
illustrated the importance of utilizing the 1-bit ACK/NAK flows in the resource
allocation. 

Note that the goodput upper bound (baseline 1) decreases with $D$ in figure
\ref{figure:numch}  because the system did not take
advantage of the frequency diversity as the selected user has to
transmit on every frequency channels. When the number of
independent channels increases, the capacity function, being
concave in channel gains, decreases.

\begin{figure}[!t]
\begin{center}
  \includegraphics[width=9cm]{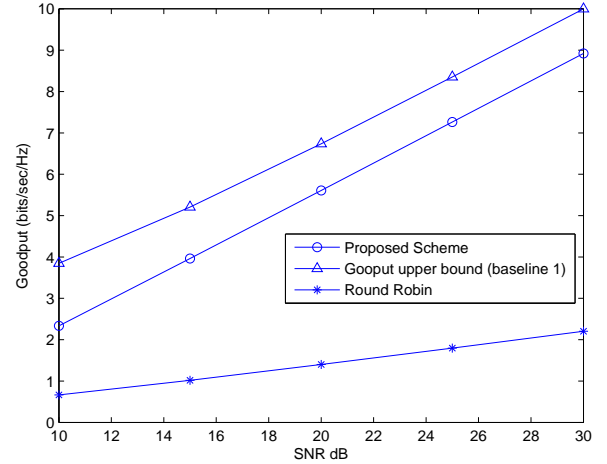}\\
  \caption{Average system goodput vs Average SNR with $P_0=24W,K=3,D=3, M=30,
PER=0.05$. The proposed solution has the same slope as the upper bound (with
perfect CSIT). }\label{figure:snr}
\end{center}
\end{figure}

\subsection{Effects of Transmit SNR}
In figure \ref{figure:snr}, there are 3 users and each user has 3
independent channels. With transmission of 30 data packets in a
time slot, the system goodput of the proposed solution achieves 60\% and 89\% of
the performance upper bound (baseline 1) in low and high SNR scenarios
respectively. Compared with baseline 2 (RR), the proposed solution has
significant 400\% gain in high SNR regime. 

\begin{figure}[!t]  
\begin{center}
  \includegraphics[width=9cm]{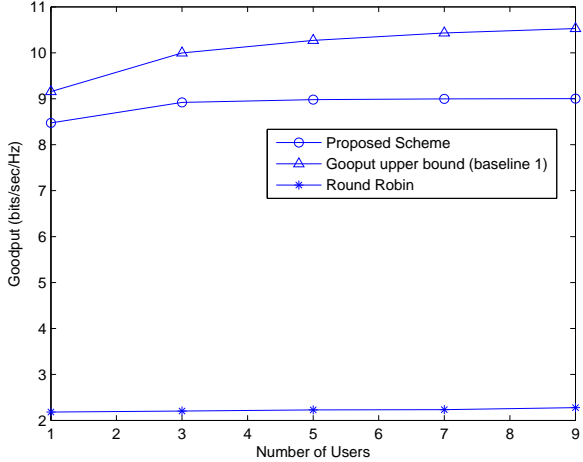}\\
  \caption{Average system
goodput vs numer of users with
transmit SNR=30dB, $P=24W, D=3, M=30, PER=0.05$.: Capacity
increases with number of users because of multi-user
diversity, so as the proposed solution.}\label{figure:user}. 
\end{center}
\end{figure}

\subsection{Effects of Number of Users}
Figure \ref{figure:user} illustrates the system goodput vs number of users for
$D=3$, $M=30, SNR = 30 dB, P_0=24W$. Similarly, the proposed scheme achieved 93
\% and 85
\% of the performance upperbound (baseline 1) with 1 user and 9 users
respectively. Compared with baseline 2 (RR), the proposed scheme achieved 400\%
goodput gain. 

\begin{figure}[!t]
\begin{center}
  \includegraphics[width=9cm]{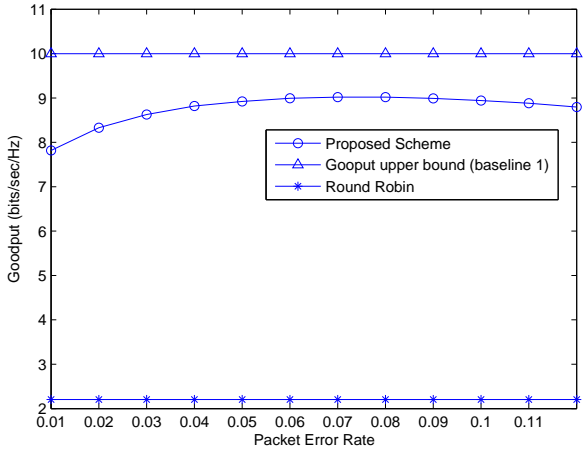}\\
  \caption{Average system goodput v.s. target PER with transmit SNR=30dB,
$K=3,D=3,M=30$: With small target PER (e.g. errors sensitive applications), the
proposed solution is
conservative and acheive a less throughput. With high
PER, the proposed solution may be
over-optimistic on channel quality. In medium PER, the proposed solution gives
the best performance.}\label{figure:outage}
\end{center}
\end{figure}

\subsection{Effects of Target PER $\epsilon$}
Figure \ref{figure:outage} illustrates the system goodput vs target PER for $SNR
=30 dB,  P_0=24W, K = 3 , M= 30$ and $D=3$. We observe that when the target PER
 is low, the proposed solution will be more conservative in
determining the transmit data rate in order to avoid packet errors due to
channel outage,
On the other hand, when the target PER is high, the proposed solution becomes
more aggressive in transmitting data but the goodput will be limited by high
channel outage probability. As a result, there is an optimal target PER, if one
is interested to optimize the system goodput.
Note that the performance upper bound of baseline 1 and the baseline 2 goodput
performance is insensitive to the target PER. 

\begin{figure}[!t]
\begin{center}
 \includegraphics[width=9cm]{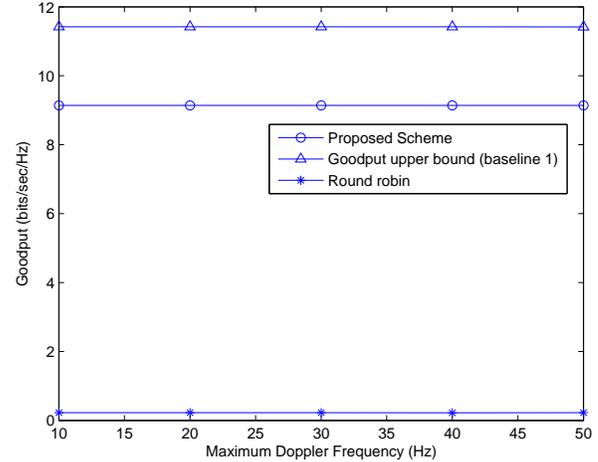}\\
\caption{Average system goodput vs maximum Doppler frequency. The users have
i.i.d. random speed (uniformly distributed from 0 to $f_{d, max}$ throughout the
simulation. $P_0=24W,K=4,D=3,M=30,SNR=30 dB, PER=0.1$ }\label{figure:doppler}
\end{center}
\end{figure}

\subsection{Effects of Mobility}
To study the robustness of the proposed scheme w.r.t. mobility, we assume the
users have i.i.d. random speed (with Doppler frequency uniformly distributed
from 0 to $f_{d,max}$). Figure \ref{figure:doppler} illustrates the average
system goodput vs $f_{d,max}$ with $SNR = 30dB, P_0=24W, K = 4$ and $D=3$.
Observe that the
proposed solution is quite robust even up to moderate mobility of 50 Hz, which
corresponds to 22.5 km/hr at 2GHz frequency. This robustness is due to the
closed-loop feedback mechanism in the proposed solution. 

 \begin{figure}[!t]
\begin{center}
  \includegraphics[width=9cm]{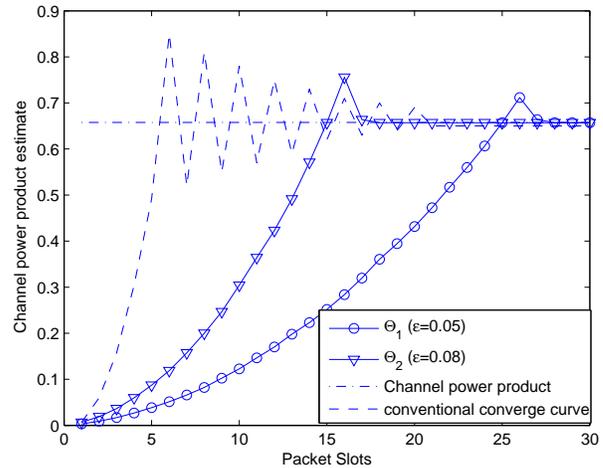}\\
  \caption{Value of channel gains estimate $\Theta$ with Different PER targets
in Different Packet Slots: The proposed solution
maximizes goodput and therefore avoids over-estimating
(resulting an NAK), hence the non-oscillating curve. A less
target PER $\epsilon$, which is more conservative, may prolong the
convergence speed.}\label{figure:theta}
\end{center}
\end{figure}

\subsection{Dynamics of Strategies}
\subsubsection{Tradeoff between Convergence Speed and Target PER}
\begin{figure}[!t]
\begin{center}
  \includegraphics[width=9cm]{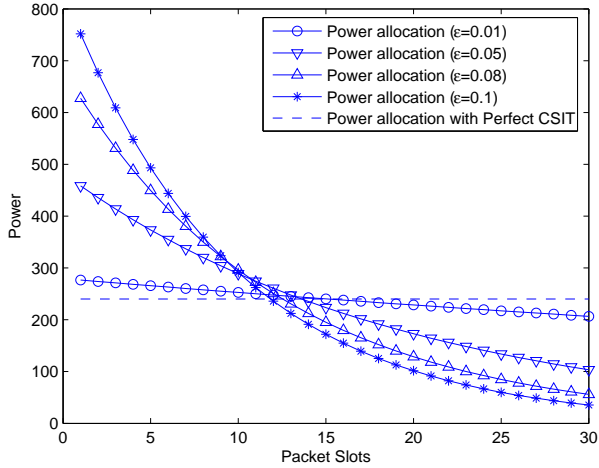}\\
\caption{Scheduled Power with Different PER Targets in
Different Packet Slots}\label{figure:power}
\end{center}
\end{figure}

\begin{figure}[!t]
\begin{center}
  \includegraphics[width=9cm]{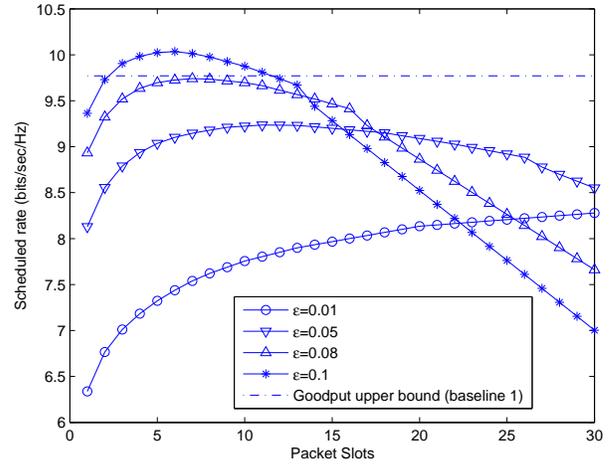}\\
  \caption{Scheduled Data Rate with Different PER Targets in Different Packet
Slots }\label{figure:rate}
\end{center}
\end{figure}

An example of the procedure of the algorithm is given in figures
\ref{figure:theta} to \ref{figure:ack_nak}. 

Figure \ref{figure:theta} plots the channel power gain estimate $\theta_m$ in a
particular channel realization v.s. time epoch $m$. ACK's are received until $m<
25$ and $m<16$ for the curves PER $\epsilon=0.5$ and 0.8 respectively. The upper
bound of the $\theta_{a_m}$ is updated with NAK and $\theta_{a_m}$ converges to
the true channel power gain product. The convergence time is shorter with high
PER. It is because large
PER provides larger flexibility for estimation.
Yet, the throughput yield from large PER may be lower
than that of small PER.

Moreover, conventional convergence curves would quickly climb
close to the channel power gain product, overshoot, oscillate and then
converge, as plotted in figure \ref{figure:theta}. The convergence
curve of our scheduling scheme would not oscillate because any
additional overshoot would waste power, time and the potential
data transmission. Thus, our scheduling scheme increases steadily, overshoots
once and converges.

\subsubsection{Power Allocation Strategies for Different Outage Target}
The power allocation of system with $ P_0=24W, K=3, D=3, M=30, SNR=30dB$,  is
plotted in figure
\ref{figure:power}. Note that the power allocation strategies
depend on the target PER $\epsilon$. The objective is to
maximize the goodput sum in all packet slots which can be
separated into current goodput and future goodput as in equation
(\ref{eqt:F_recursion_short}). 
To maximize the goodput sum for large PER, more power should be allocated at the
early slots to have as much successful transmission as possible
. Notice that, as PER
decreases, the power allocation converges to the power allocation
for perfect CSIT, equal power allocation. It is because at the
extreme case of zero PER, the probability of getting
outage is zero, meaning that we have perfect CSIT (baseline 1).

\subsubsection{Rate Allocation Strategies for Different PER
Target} Assume $P_0=24W,SNR=30dB, D=3, K=3, M=30$. The rate allocation curves
with different PER
target
are plotted in figure \ref{figure:rate}. Note that the area under
the curve is the throughput. The data rate achieved by baseline 1 is plotted
with a dotted line. Notice that the area achieved
by small PER, 0.01, is small and the area increases by
increasing the PER. However, area decreases after PER
0.07 which is the optimal PER in the current system assumption.
An over-conservative PER target would yield too little
goodput as the $\mathbb{X}_{a_m}$ is under estimated. An
over-optimistic PER target would also yield a low goodput as
outage occurs when $\mathbb{X}_{a_m}$ is over estimated.

The allocated rate $r_m$ increases with the increment of knowledge of
the channel power gain in figure \ref{figure:rate}. Then $r_m$ decreases after
slot 10 because the scheduler has spent half of the total power in the
first 10 slots. Less rate is resulted from smaller power remained for these 20
slots. 

\begin{figure}[!t]
\begin{center}
  \includegraphics[width=9cm]{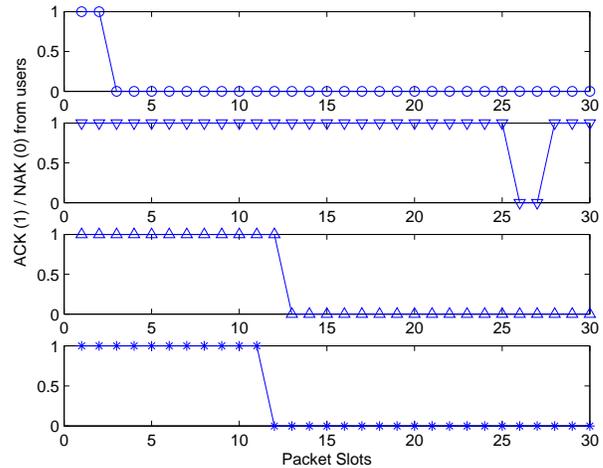}\\
  \caption{Acknowledgements from Different Users, top (user 1), second one from
the top (user 2) and so on}\label{figure:ack_nak}
\end{center}
\end{figure}

\subsubsection{Acknowledgements Reveal CSIT}
In figure \ref{figure:ack_nak}, the acknowledgements from user 1
(from the top) to user 4 (from the bottom) are plotted whereas 1
denotes positive acknowledgement (ACK) and 0 denotes negative
acknowledgement (NAK). After each transmission, each user 
decodes the packet header and feedback to transmitter. If a user $k$ reports
NAK at slot $m$, user $k$ would have a channel power gain less than the
channel power gain estimate at slot $m$, $\theta_m$. Thus, we know that
$\theta_2 \leq X_1 < \theta_3$, $\theta_{11} \leq X_4 <
\theta_{12} \leq X_3 < \theta_{13}$. Since NAK are received at
slot 25 and 26, we know that $\theta_{24} \leq X_2 < \theta_{27}$.

\section{Conclusions}\label{section:conclusion}
In this paper, we considered the OFDM resource optimization problem based on
ACK/NAK feedbacks from the mobiles without explicit CSIT at the base station. We
derive a simple closed-form solution for the MDP cross-layer problem which is
asymptotically optimal for sufficiently small target PER. The proposed solution
also has low complexity and is suitable for realtime implementation. Simulation
results revealed that the system goodput performance of the proposed solution
achieved 89\% of the performance upper bound (perfect CSIT performance) and has
over 400\% gain compared to round robin scheduling. Due to the built in
closed-loop feedback mechanism, the proposed scheme is shown to have robust
performance against CSIT errors and different mobility. Asymptotic analysis is
also provided to obtain useful design insights. 

\section{Appendix}
\subsection{Recursive Property of Goodput} \label{Appendix:resur_goodput}
Recall from equation (\ref{eqt:opt_problem}). Expectation over the channel power
$H$ is the same as the iterative expectation $E_{\hat{V}_m} E_{H|\hat{V}_m}$
where $\hat{V}_m$ is the feedbacks from users from slot $m$ to $M$. 
 Recall $V_m$, defined in \eqref{eqt:causal_feedbacks} is the causal feedbacks
from slot 1 to $m-1$. Combining $V_m$ and $\hat{V}_m$ gives the whole history:
$(V_m,\hat{V}_m)=V_M$.

\begin{equation}
F_m(\bar{P}_m,W_{m-1})=\underset{\mathbf{p}_m, \mathbf{r}_m, \mathbf{a}_m}{max}
E_{\hat{V}_m} E_{H|\hat{V}_m}\left\{ \sum_{i=m}^M v_{a_i,i} r_i\right\}.
\end{equation}
Evaluating the expectation yields
\begin{equation}\label{eqt:F_recur}
F_m(\bar{P}_m,W_{m-1})=\underset{\mathbf{p}_m, \mathbf{r}_m, \mathbf{a}_m}{max}
E_{\hat{V}_m} \left\{  \sum_{i=m}^M Pr(c(p_i, X_{a_i})>r_i) r_i\right\}.
\end{equation}

Separate the instantaneous goodput at slot $m$ from the goodput sum from slot
 $m+1$ to $M$. Take an iterative expectation and obtain equation \eqref{eqt:F1}.
\begin{figure*}[!t]
\normalsize
\setcounter{MYtempeqncnt}{\value{equation}}
\setcounter{equation}{62}
\begin{equation}\label{eqt:F1}
F_m(\bar{P}_m, W_{m-1})= \underset{\mathbf{p}_m,\mathbf{r}_m,\mathbf{a}_m}{max}
E_{\mathbf{v}_m} E_{\hat{V}_{m+1}|\mathbf{v}_m} \left\{Pr(c(p_m, X_{a_m})>r_m)
r_m+ \sum_{i=m+1}^M Pr(c(p_i, X_{a_i})>r_i) r_i\right\}.
\end{equation}

\begin{equation}\label{eqt:F2}
F_m(\bar{P}_m, W_{m-1})= \underset{\mathbf{p}_m,\mathbf{r}_m,\mathbf{a}_m}{max}
\left\{ Pr(c(p_m, X_{a_m}>r_m)|W_{m-1}) r_m + E_{\mathbf{v}_m}  \left\{
E_{\hat{V}_{m+1}|\mathbf{v}_m} \sum_{i=m+1}^M
Pr(c(p_i, X_{a_i})>r_i|W_{i-1}) r_i\right\} \right\}
\end{equation}

\begin{equation}\label{eqt:F3}
F_m(\bar{P}_m, W_{m-1})= \underset{\mathbf{p}_m,\mathbf{r}_m,\mathbf{a}_m}{max}
\left\{
Pr(c(p_m, X_{a_m})>r_m|W_{m-1}) r_m+ E_{\mathbf{v}_m}  F_{m+1}(\bar{P}_m-p_m,
W_m)
\right\}.
\end{equation}

\begin{equation}\label{eqt:F_in}
F_m(\bar{P}_m,W_{m-1})=\underset{\underset{p_m,\ldots,p_M,r_m,\ldots,r_M}{a_m,
\ldots,a_M}}{max}
\left\{ (1-\epsilon)\frac{NT}{DM}\log_2 \left( \left( \frac{p_m}{N}
\right)^D \theta_m \right)+ (1-\epsilon) F_{m+1}(\bar{P}_m-p_m,W_m |
v_m=1) \right\}
\end{equation}
\begin{equation}\label{eqt:F_in_long}
F_m(\bar{P}_m,W_{m-1})=
\underset{\underset{p_m,\ldots,p_M,r_m,\ldots,r_M}{a_m,\ldots,a_M}}{max}
\left\{ (1-\epsilon)\frac{NT}{DM}\log_2 \left( \left( \frac{p_m}{N}
\right)^D \theta_m \right)+ \cdots + (1-\epsilon)^{M-m+1}\frac{NT}{DM}\log_2
\left( \left( \frac{p_M}{N} \right)^D \theta_M \right) \right\}
\end{equation}
\begin{equation}\label{eqt:F_long}
F^{(1)}_m(\bar{P}_m,W_{m-1})=
\underset{p_m,\ldots,p_M,r_m,\ldots,r_M}{max} \left\{
\frac{NT}{M}(1-\epsilon)\left[ \log_2\left( \frac{p_m}{N} \right)+ \cdots+
(1-\epsilon)^{M-m} \log_2 \left(\frac{p_M}{N}
\right)\right] \right.
\end{equation}
\begin{equation*}
\left. \frac{NT}{DM}(1-\epsilon) \left[ \log_2(\theta_m) + \cdots
+ \left( 1- \epsilon \right)^{M-m}\log_2\left(
\theta_M\right)\right] \right\}
\end{equation*}
\setcounter{equation}{\value{MYtempeqncnt}}
\hrulefill
\vspace*{4pt}
\end{figure*}

Since the first term does not depend on $V_m$ nor $v_m$, it simplifies to
\eqref{eqt:F2}.

Note that the second term is the expectation of $F_{m+1}(\bar{P}_m-p_m, W_{m})$
over $\mathbf{v}_m$ according to equation (\ref{eqt:F_recur}). Equation
\eqref{eqt:F3} can be obtained.

\subsection{Dynamics of $\theta_m$}\label{appendix:theta}
Denote the event $X_k \geq L(\Theta_{m-1}, \bar{v}_k^{m-1})$ by
$\mathcal{L}$ and $X_k < U(\Theta_{m-1}, \bar{v}_k^{m-1})$ by
$\mathcal{U}$ respectively. Employ the theorem of conditional
probability on equation \eqref{eqt:quality_theta}.
\begin{equation}\label{eqt:cond_pr}
\frac{Pr(X_k \geq \theta_m, \mathcal{L},\mathcal{U}
)}{Pr(\mathcal{L},\mathcal{U})}=1-\epsilon
\end{equation}
Recall the cdf of $X_k$, $\phi$, in \eqref{eqt:phi}, 
\eqref{eqt:cond_pr} can
be rewritten as
\begin{equation}
\frac{\phi(U(\Theta_{m-1},
\bar{v}_k^{m-1}))-\phi(\theta_m)}{\phi(U(\Theta_{m-1},
\bar{v}_k^{m-1}))-\phi(L(\Theta_{m-1},
\bar{v}_k^{m-1}))}=1-\epsilon
\end{equation}
Rearranging the terms and equation (\ref{eqt:quality_theta}) can
be obtained.

\subsection{Optimal User Selection}\label{Appendix:opt_user}
This section is to prove that the user selection
$a_m=\underset{k}{\arg\max} L(\Theta_{m-1},\bar{v}_k^{m-1})$ 
maximizes
$F_m(\bar{P}_m,W_{m-1})$ in \eqref{eqt:F_recursion_short}. Substitute
$\theta_m= \left( \frac{N}{p_m} \right)^D 2^{\frac{DM r_m}{NT}}$ to
$F_m(\bar{P}_m,W_{m-1})$ and we obtain equation
\eqref{eqt:F_in}. 

Further expand \eqref{eqt:F_in}, we obtatin \eqref{eqt:F_in_long}

 As we assume $v_m,\ldots,v_M=1$, we have
$\theta_m=L(\Theta_m, \bar{v}_{a_{m+1}}^m)$ and therefore

\begin{equation}\label{eqt:phi_sub}
\theta_{m+1}=\phi^{-1} \left( \epsilon \phi(U(\Theta_m,
\bar{v}^m_{a_{m+1}}))+ (1-\epsilon)\phi(\theta_m)\right).
\end{equation}
As $\phi(\theta_m)$ is the CDF of $\theta_m$, $\phi(\theta_m)$ is
monotonic increasing with $\theta_m$, so as $\phi^{-1}$. Thus,
$\theta_{m+1}$ increases with $\theta_m$. According to equation
(\ref{eqt:F_in_long}), $F_m(\bar{P}_m,W_{m-1})$ increases with $\theta_m$.
What remains to prove is that $a_m=\underset{k}{\arg\max}\;
L(\Theta_{m-1},\bar{v}_k^{m-1})$ maximizes $\theta_m$. We prove by
contradiction. Let $k^*
\neq a_m$, we have $L(\Theta_{m-1},\bar{v}_{k^*}^{m-1}) <
L(\Theta_{m-1},\bar{v}_{a_m}^{m-1})$ by definition, and
$U(\Theta_{m-1},\bar{v}_{k^*}^{m-1}) \leq
L(\Theta_{m-1},\bar{v}_{a_m}^{m-1}) \leq
U(\Theta_{m-1},\bar{v}_{a_m}^{m-1})$ by characteristics. Denote
$\theta_m$ by $\Psi(k)$ where $k$ is the user selection in slot
$m$. According to equation (\ref{eqt:phi_sub}), $\Psi(k)<\Psi(a_m)
\; \forall k \neq a_m$. Therefore, $a_m=\underset{k}{\arg\max}\;
L(\Theta_{m-1},\bar{v}_k^{m-1})$ maximizes $\theta_m$ and
therefore $F_m(\bar{P}_m,W_{m-1})$.

\subsection{Optimal Power selection}\label{Appendix:opt_power}
At the base case, we would like to
maximize the goodput in the last slot $M$ which is to solve

\begin{equation}
F^{(1)}_M(\bar{P}_M,W_{M-1})= \underset{p_M,r_M}{max} (1-\epsilon)r_M.
\end{equation}
And given $W_{M-1}$, $\theta_m$ can be solved by taking an inverse
of the function $\phi_{a_m}(.)$ in equation
(\ref{eqt:theta_dynamics})
\begin{equation}\label{eqt:theta_dynamics_inverse}
\theta_m=\phi^{-1} \left(\epsilon \phi
(U(\Theta_{m-1},\bar{v}_{a_m}^{m-1}))+(1- \epsilon)\phi
(L(\Theta_{m-1},\bar{v}_{a_m}^{m-1})) \right)
\end{equation}
As the relation of power and rate is $\theta_M=\left(
\frac{N}{p_M}\right)^D 2^{\frac{DM r_M}{NT}}$, the optimal solution
at the base case is
\begin{equation}
\left\{%
\begin{array}{ll}
    p_M=\bar{P}_M \\
    r_M=\frac{NT}{M} \log_2 \left( \frac{\bar{P}_M}{N} \right) +
\frac{NT}{DM} \log_2(\theta_M) \\
\end{array}%
\right.
\end{equation}
Therefore, $r_m$ can be solely expressed by $\theta_{m}$ and
$p_m$. Recursively develop $F_1(P_0)$, we have
\begin{equation}
F^{(1)}_1(P_0,W_0)= \underset{p_1, r_1}{max} \left\{
(1-\epsilon) r_1 + \cdots + (1-\epsilon)^M r_M\right\}
\end{equation}
With some mathematic manupulation, we obtain equation \eqref{eqt:F_long}.

As we have assumed $v_m=1$, $\theta_m$ can be computed for $m=1$
to $M$. Note that $p_{m+1},\ldots,r_M$ are of the form
\begin{eqnarray}\label{eqt:Ps}
\nonumber p_{m+1} &=& a_1(\bar{P}_m-p_m)\\
p_{m+2} &=& a_2(\bar{P}_m-p_m-p_{m+1})\\
\nonumber & & \vdots\\
\nonumber p_M &=& a_{M-m}(1-a_{M-m-1})\cdots (1-a_1)(\bar{P}_m-p_m)
\end{eqnarray}
Therefore, the closed form of optimal power allocation is
obtained. Note that the objective function in (\ref{eqt:F_long})
is concave in $p_m$. Substitute equation (\ref{eqt:Ps}) to
$F^{(1)}_m(\bar{P}_m,W_{m-1})$ in equation (\ref{eqt:F_long}) and
differentiate
it and set it to zero. We obtain
\addtocounter{equation}{6}
\begin{equation}\label{eqt:power_alloc}
p_m= \frac{\epsilon \bar{P}_m}{1- (1-\epsilon)^{M-m+1}}
\end{equation}
which is solely depending on $\epsilon$ and $\bar{P}_m$ but nothing
else. The
solutions obtained here is a lower bound of the original solution
as the objective is solving the problem in only one direction
which assumes all positive feedbacks and correspond to the all
positive routes in the decision tree.

\subsection{Shrinking User Selection Set
$\mathbb{K}_m$}\label{Appendix:shrink_user_set}
Before proving this lemma, we need to introduce two properties of the lower
bound of channel power $X_k$, $L(\Theta_m,\bar{v}_k^m)$.
\subsubsection{Monotonic Increasing Lower Bound of Real Channel
Power}\label{Appendix:mono_lower_bound}
\begin{Lem}\label{Lem:monotonic_lower_bound}
The lower bound of the channel power gains $L(\Theta_m, \bar{v}_k^m)$ increases
monotonically  with $m$ .
\end{Lem}

\begin{proof}
{\small
\begin{eqnarray}\label{eqt:monotonic_lower_bound}
& &L(\Theta_m, \bar{v}_k^m) \\
\nonumber &=& \max \left\{ \theta_i : v_{k,i}=1, 1\leq i \leq
m
\right\}\\
\nonumber &=& \begin{cases}
     \max \left\{ \theta_m , \left\{ \theta_i : v_{k,i}=1, 1 \leq i \leq
m-1\right\} \right\} & \text{if $v_{k,m}=1,$}  \\
     \max \left\{ \theta_i : v_{k,i}=1, 1 \leq i \leq m-1 \right\} & \text{if
$v_{k,m}=0.$}
\end{cases} \\
\nonumber &=& \begin{cases}
     \max \left\{ \theta_m, L(\Theta_{m-1}, \bar{v}_k^{m-1}) \right\} & \text{if
$v_{k,m}=1,$}  \\
     L(\Theta_{m-1}, \bar{v}_k^{m-1}) & \text{if $v_{k,m}=0.$}
\end{cases} \\
\nonumber &\geq& L(\Theta_{m-1}, \bar{v}_k^{m-1})
\end{eqnarray}}
\end{proof}

\subsubsection{Lower Bound of Channel Power of Selected User Larger than the
upper bound of channel power of the Remaining
Users}\label{Appendix:up_low_bound}
\begin{Lem}\label{Lem:Upper_Lower_Bound}
Assume $\exists k \not\in \mathbb{K}_{m-1}$.

\begin{equation}
U(\Theta_{m-1},\bar{v}_k^{m-1}) \leq L(\Theta_{m-1},\bar{v}_{k'}^{m-1}) \;
\forall k' \in \mathbb{K}_{m-1}
\end{equation}
\end{Lem}

\begin{proof}
Assume $\exists k \not\in \mathbb{K}_{m-1}$.
Recall equation (\ref{eqt:bounds_update}), 
\begin{equation*}
U(\Theta_{m-1},\bar{v}_k^{m-1})= \min \left\{ \theta_i : v_{k,i} =0 , 1 \leq i
\leq m-1 \right\}
\end{equation*}
There exist a packet slot $q$, $1 \leq q \leq m-1$, such that
$v_{k,q}=0$ and $v_{k',q}=1$, which can be described mathematically in equation
\eqref{eqt:ineq}.

\begin{figure*}[!t]
\normalsize
\setcounter{MYtempeqncnt}{\value{equation}}
\setcounter{equation}{75}
\begin{eqnarray}
U(\Theta_{m-1}, \bar{v}_k^{m-1})&=& \min \left\{ \theta_q, \left\{ \theta_i:
v_{k,i}=0, 1\leq i \leq q-1, q+1 \leq i \leq m-1 \right\} \right\} \\
\nonumber &=& \min \left\{ \theta_q, U(\Theta_{q-1},\bar{v}_k^{q-1}),
U(\Theta_{m-1},\bar{v}_k^{m-1})\right\} \label{eqt:ineq}
\end{eqnarray}
\setcounter{equation}{\value{MYtempeqncnt}}
\hrulefill
\vspace*{4pt}
\end{figure*}

From definition, $\theta_q \geq L(\Theta_{q-1}, v_{k'}^{q-1})$ and $L(\Theta_q,
v_{k'}^q)= \max \left\{ \theta_q, L(\Theta_{q-1}, v_{k'}^{q-1}) |
v_{k',q}=1\right\}$. Thus, $L(\Theta_q, v_{k'}^q)= \theta_q$ if $v_{k',q}=1$.
Thus, continuing from equation (\ref{eqt:ineq})
\begin{eqnarray}
& &U(\Theta_{m-1}, \bar{v}_k^{m-1})\\
\nonumber &=& \min \left\{L(\Theta_{q},\bar{v}_{k'}^q),
U(\Theta_{q-1},\bar{v}_{k}^{q-1}), U(\Theta_{m-1},\bar{v}_k^{m-1}) \right\}\\
\nonumber &=& L(\Theta_{q},\bar{v}_{k'}^q)\\
\nonumber &\leq& L(\Theta_{m-1},\bar{v}_{k'}^{m-1})
\end{eqnarray}
The last inequality is proved by lemma \ref{Lem:monotonic_lower_bound}.
\end{proof}

We are going to prove this lemma by contradiction. 
Assume $\exists k \in \mathbb{K}_m$ and $ k \not\in \mathbb{K}_{m-1}$. At slot
$m$, $\forall k' \in \mathbb{K}_{m-1}, k' \not\in \mathbb{K}_m$, by lemma 
\ref{Lem:monotonic_lower_bound}, the lower bound of channel power gain is
monotonically increasing with $m$. 
\begin{equation}
L(\Theta_{m}, \bar{v}_{k'}^{m})\geq L(\Theta_{m-1}, \bar{v}_{k'}^{m-1})
\end{equation}
Also, by lemma \ref{Lem:Upper_Lower_Bound}, all users outside the user selection
set have upper bound less than or equal to that of users inside the user
selection set. $\forall k \not\in \mathbb{K}_{m-1}, k' \in \mathbb{K}_{m-1}$
\begin{equation}
U(\Theta_{m-1},\bar{v}_k^{m-1})\leq L(\Theta_{m-1}, \bar{v}_{k'}^{m-1})
\end{equation}

Because $k \in \mathbb{K}_m, k' \not\in \mathbb{K}_m$, we have
\begin{equation}
L(\Theta_m, \bar{v}_k^m)>L(\Theta_m, \bar{v}_{k'}^m).
\end{equation}
Thus, we have
\addtocounter{equation}{1}
\begin{eqnarray}
& & L(\Theta_m, \bar{v}_k^m)\\
\nonumber &>& L(\Theta_m,\bar{v}_{k'}^m) \hspace{1cm} (\forall k
\in \mathbb{K}_m, k' \not\in \mathbb{K}_m)\\
\nonumber &\geq& L(\Theta_{m-1},\bar{v}_{k'}^{m-1}) \hspace{0.5cm} (\mbox{by
lemma 7})\\
\nonumber &\geq& U(\Theta_{m-1},\bar{v}_{k}^{m-1}) \hspace{0.5cm} (\forall k'
\in \mathbb{K}_{m-1}, k \not\in \mathbb{K}_{m-1})
\end{eqnarray}
which leads to a contradiction. Thus, $\forall k \in \mathbb{K}_m, k \in
\mathbb{K}_{m-1}$.

\subsection{Channel Estimate of Selected User between Upper and Lower
Bound}\label{Appendix:user_ch_bound}
We are going to prove this claim by mathematical induction. In the base case,
$m=0$, before any transmission, we have initialization
\begin{eqnarray}
L&=& 0 \\
U &=& \infty\\
X_k &\in& \left[ L, U \right] \; \; \forall k \in \mathbb{K}_0
\end{eqnarray}
where $\mathbb{K}_0=\left\{ 1, \ldots, K \right\}$.

Assume the statement is true for $m=q$. We obtain
\begin{equation}
X_k \in \left[ L(\Theta_q, \bar{v}_k^q), U(\Theta_q,\bar{v}_k^q)\right], \;
\forall k \in \mathbb{K}_q
\end{equation}
When $m=q+1$, before the $(q+1)$-th transmission,
\begin{equation}
L(\Theta_q, \bar{v}_k^q) \leq \theta_{q+1} \leq U(\Theta_q, \bar{v}_k^q), \;
\forall k \in \mathbb{K}_q
\end{equation}
After $(q+1)$-th transmission, there are two cases, either ACK or NAK. If an ACK
is received then we have
\begin{eqnarray}
& & r_{q+1} \leq c(p_{q+1}, X_k)\\
\nonumber & \mbox{or} & \frac{NT}{DM} \log_2 \left( \left(
\frac{p_{q+1}}{N}\right)^D
\theta_{q+1}\right) \\
\nonumber && \hspace{2cm} \leq \frac{NT}{DM} \log_2 \left(
\left(\frac{p_{q+1}}{N}\right)^D X_k \right)\\
\nonumber & \mbox{or} & \theta_{q+1} \leq X_k.
\end{eqnarray}
The updates of the bounds are
\begin{eqnarray}
L(\Theta_{q+1},\bar{v}_k^{q+1}) &=& \max \left\{ L(\Theta_{q},\bar{v}_k^q),
\theta_{q+1} \right\}\\
\nonumber &=& \theta_{q+1}\\
\mbox{and } U(\Theta_{q+1}, \bar{v}_k^{q+1}) &=& U(\Theta_{q}, \bar{v}_k^{q}).  
\end{eqnarray}
Thus, we have $\forall k \in \mathbb{K}_q \bigcap \left\{ k:
v_{k,q+1}=1\right\}$
\begin{equation}
L(\Theta_{q+1}, \bar{v}_k^{q+1}) \leq X_k \leq U(\Theta_{q+1},\bar{v}_k^{q+1}).
\end{equation}
Let $\mathbb{K}_{q+1}= \mathbb{K}_q \bigcap \left\{ k: v_{k,q+1}=1 \right\}$
which completes the proof.
Similarly, if NAK is received, $X_k \leq \theta_{q+1}$. The updates of the
bounds are
\begin{eqnarray}
L(\Theta_{q+1},\bar{v}_k^{q+1})&=&L(\Theta_q,\bar{v}_k^q)\\
U(\Theta_{q+1},\bar{v}_k^{q+1})&=& \theta_{q+1}
\end{eqnarray}
Thus, we have $\forall k \in \mathbb{K}_q \bigcap \left\{ k:
v_{k,q+1}=0\right\}$
\begin{equation}
L(\Theta_{q+1}, \bar{v}_k^{q+1}) \leq X_k \leq U(\Theta_{q+1},\bar{v}_k^{q+1}).
\end{equation}
Let $\mathbb{K}_{q+1}= \mathbb{K}_q \bigcap \left\{ k: v_{k,q+1}=0 \right\}$
which completes the proof.

\subsection{Monotonic Decreasing Gap between Upper and Lower
Bounds}\label{Appendix:bound_gap}
The difference of the gaps at slot $m$ and $m-1$ is
\begin{eqnarray}
& & w_m-w_{m-1} \\
\nonumber &=& \left\{ U(\Theta_m, \bar{v}_{a_m}^m)-L(\Theta_m,
\bar{v}_{a_m}^m)\right\} \\
\nonumber & & - \left\{U(\Theta_{m-1},
\bar{v}_{a_{m-1}}^{m-1})-L(\Theta_{m-1}, \bar{v}_{a_{m-1}}^{m-1}) \right\}\\
\nonumber &=& \begin{cases}
     L(\Theta_{m-1}, \bar{v}_{a_{m-1}}^{m-1})-L(\Theta_m, \bar{v}_{a_m}^m) &
\text{if  } \bar{v}_{a_m}^m=1, \\
     U(\Theta_m, \bar{v}_{a_m}^m)-U(\Theta_{m-1}, \bar{v}_{a_{m-1}}^{m-1}) &
\text{if } \bar{v}_{a_m}^m=0.
\end{cases}\\
\nonumber &=&  \begin{cases}
     L(\Theta_{m-1}, \bar{v}_{a_{m-1}}^{m-1})-\theta_m & \text{if  }
\bar{v}_{a_m}^m=1, \\
     \theta_m-U(\Theta_{m-1}, \bar{v}_{a_{m-1}}^{m-1}) & \text{if }
\bar{v}_{a_m}^m=0.
\end{cases}\\
\nonumber &\leq& 0 \; \forall m
\end{eqnarray}
The last inequality is due to the fact that
$L(\Theta_{m-1},\bar{v}_{a_{m-1}}^{m-1}) \leq \theta_m \leq
U(\Theta_{m-1},\bar{v}_{a_{m-1}}^{m-1})$

\subsection{Scheduled Rate Achieves Capacity}\label{Appendix:rate_cap}
By lemma \ref{lem:subset}, when $m \rightarrow \infty$, the user selection set
degenerates to a single user whose has the largest lower bound of the channel
power gains, $\mathbb{K}_m= k$ where $L(\Theta_m, \bar{v}_k^m) > L(\Theta_m,
\bar{v}_{k'}^m)$ and $k \neq k'$. Using lemma \ref{lem:real_in_bounds} and
\ref{thm:gap_decrease}, we have
\begin{equation}
m \rightarrow \infty, \; L(\theta_m, \bar{v}_{a_m}^m)=U(\theta_m,
\bar{v}_{a_m}^m)=X_{a_m}
\end{equation} 
Thus, we have
\begin{equation}
m \rightarrow \infty, \;  \mathbb{K}_m=a_m=k, \mbox{    where } X_k > X_{k'}, \;
\; k \neq k'
\end{equation}
Also by lemma \ref{thm:gap_decrease}, we have
\begin{equation}
m \rightarrow \infty, \; \theta_m=L(\theta_m, \bar{v}_{a_m}^m)=U(\theta_m,
\bar{v}_{a_m}^m)=X_{a_m}
\end{equation} 
Thus, we have the scheduled rate at slot $m$,
\begin{eqnarray}
\underset{m \rightarrow \infty}{\lim}r_m &=& \underset{m \rightarrow
\infty}{\lim} \frac{NT}{DM} log_2 \left( \left( \frac{p_m}{N}\right)^D
\theta_{m}\right)\\
\nonumber &=& \underset{m \rightarrow \infty}{\lim} \frac{NT}{DM} log_2 \left(
\left(\frac{p_m}{N}\right)^D X_{k}\right)\\
\nonumber &=& c(p_m, X_k)
\end{eqnarray} where user $k$ has the largest channel power gains. The quantity
$c(p_m, X_k)$ is the capacity achieved by the system with perfect CSIT.

\begin{biography}[{\includegraphics[width=1in,height=1.25in,clip,
keepaspectratio]{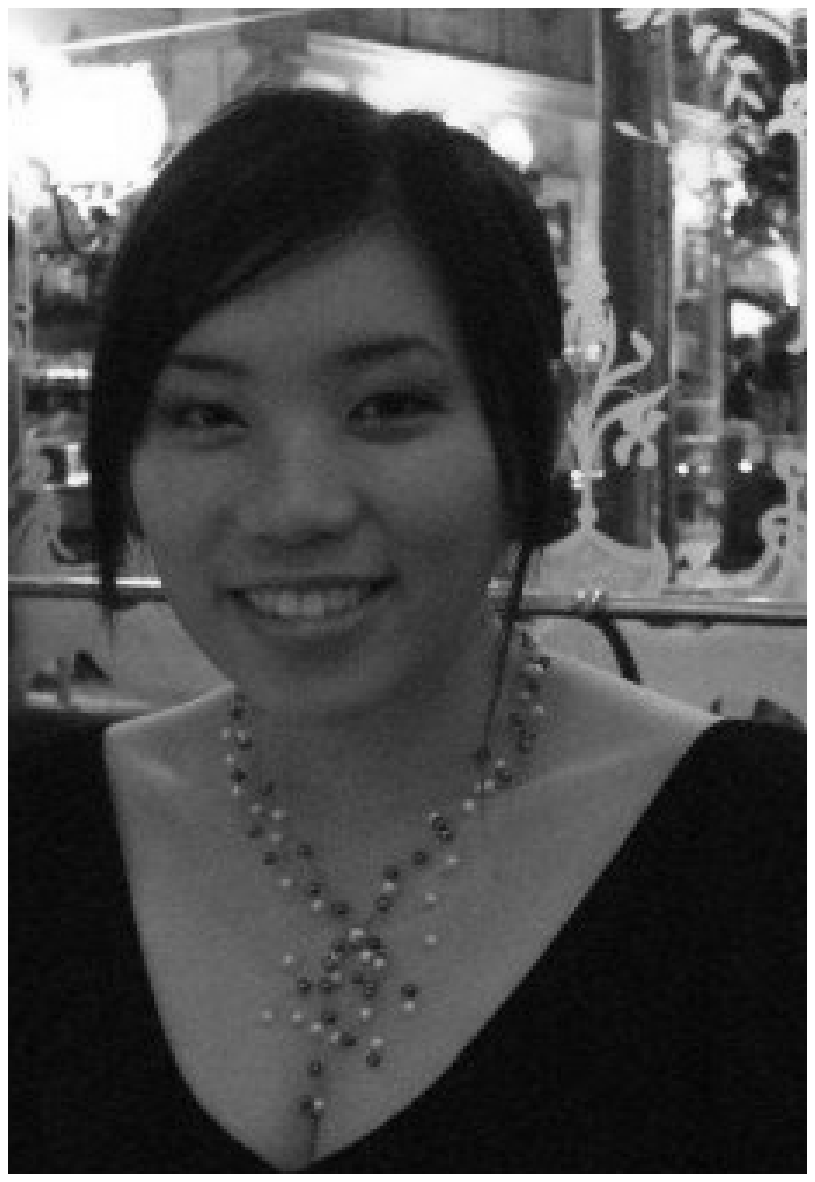}}]{Zuleita K.~M.~Ho (S'05)} enrolled into Hong Kong
University of Science and Technology
through the Early Admission for Outstanding Secondary Six Students in 2001and 
graduated from the Dept of ECE, with a B.Eng (Distinction 1st Hons)  and
M.Phil
in 2004 and 2006. She is currently a Ph.D candidate in mobile communications
in
EURECOM, France.
Her research interests include cooperations in wireless networks, distributed
barginaing, Game theory, crosslayer optimization, information theory and
random
matrices. 
Zuleita has received more than 10 academic scholarships including, The
Croucher Foundation Scholarship 2007, The Hongkong Bank
Foundation Overseas Scholarship Schemes 2004, which sponsors a full year study
at Massachusetts Institute of Technology and The IEE Outstanding Student Award
2004.
\end{biography}

\begin{biography}
[{\includegraphics[width=1in,height=1.25in,clip,keepaspectratio]{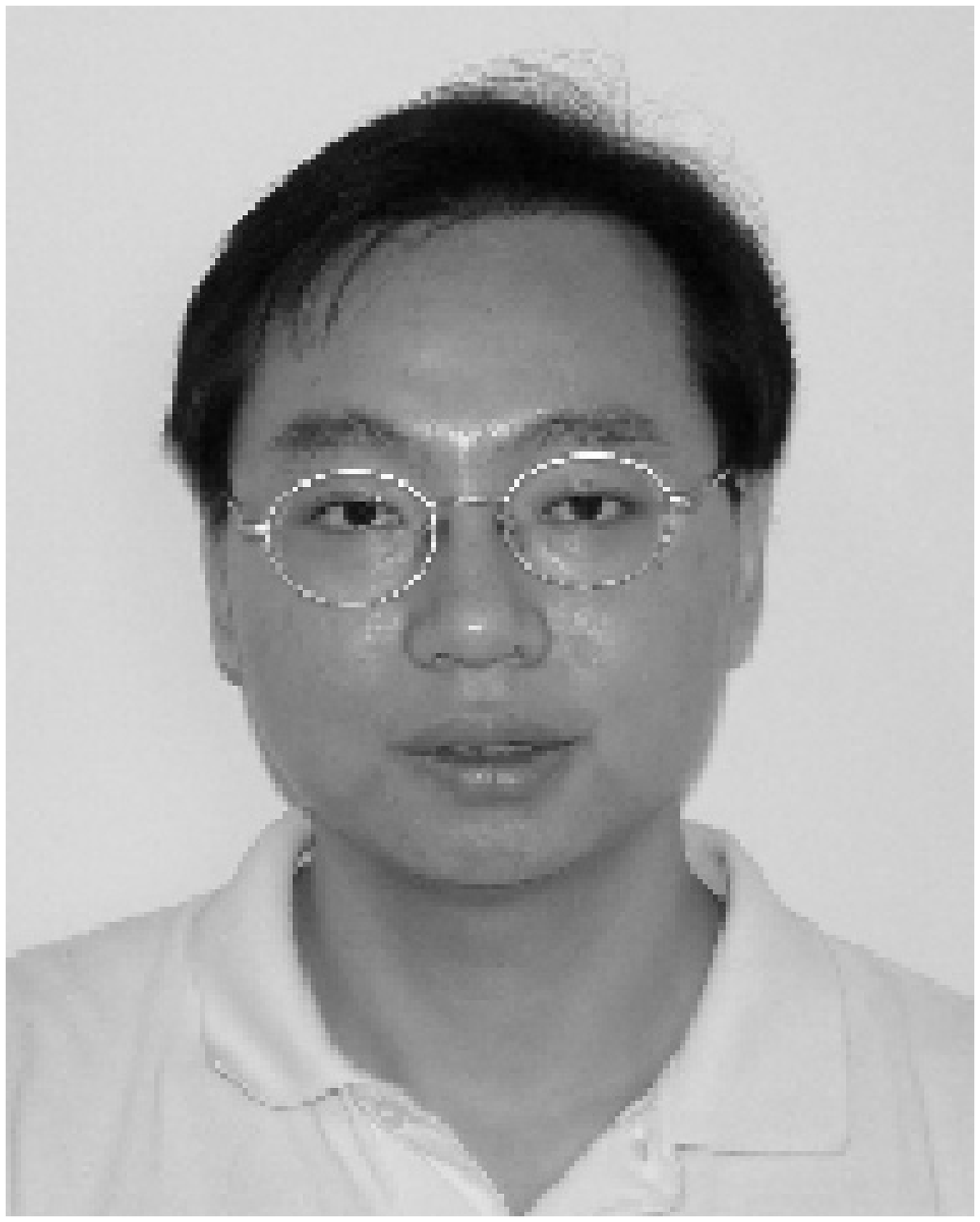}}]{
Vincent K.N. Lau (M'97- SM'01)}
obtained B.Eng (Distinction 1st Hons) from the University of Hong Kong
(1989-1992) and Ph.D. from Cambridge University (1995-1997). He was with HK
Telecom (PCCW) as system engineer from 1992-1995 and Bell Labs - Lucent
Technologies (NJ) as member of technical staff from 1997-2003. He then joined
the Department of ECE, Hong Kong University of Science and Technology (HKUST) as
Associate Professor. His current research interests include the 
robust and
delay-sensitive cross-layer scheduling of MIMO/OFDM wireless systems with
imperfect channel state information, cooperative and cognitive communications,
dynamic spectrum access as well as stochastic approximation and Markov Decision
Process. He served as the editor of IEEE Transactions on Wireless
Communications, guest editor of IEEE Journal on Selected Areas in Communications
(JSAC), IEEE Journal of Special Topics on Signal Processing, IEEE System Journal
as well as EURASIP Journal on Wireless Communications and Networking. 
\end{biography}

\begin{biography}[{\includegraphics[width=1in,height=1.25in,clip,
keepaspectratio
]{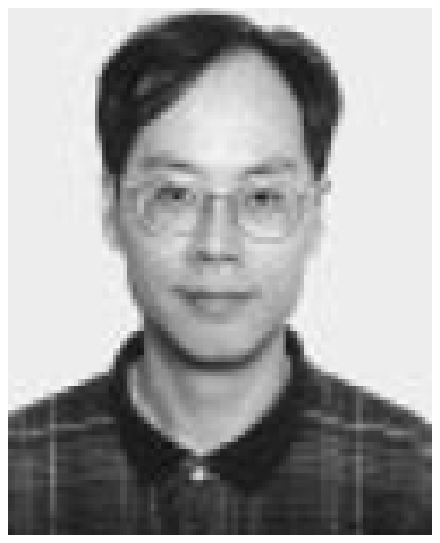}}]{
Roger S.-K. Cheng (S'86-M'92)} received the B.S.
degree from Drexel University, Philadelphia, PA, in
1987, and the M.A. and Ph.D. degrees from Princeton
University, Princeton, NJ, in 1988 and 1991, respectively,
all in electrical engineering.
From 1987 to 1991, he was a Research Assistant in
the Department of Electrical Engineering, Princeton
University, Princeton, NJ. From 1991 to 1995, he was
an Assistant Professor in the Electrical and Computer
Engineering Department, University of Colorado at
Boulder. In June 1995, he joined the Faculty of the
Hong Kong University of Science and Technology, where he is currently an
Associate
Professor in the Department of Electrical and Electronic Engineering. He
has also held visiting positions with Qualcomm, San Diego, CA, in the summer
of 1995, and with the Institute for Telecommunication Sciences, NTIA, Boulder,
CO, in the summers of 1993 and 1994. His current research interests include
wireless communications, OFDM, space?time processing, CDMA, digital
implementation
of communication systems, wireless multimedia communications,
information theory, and coding.
Dr. Cheng is currently an Editor for Wireless Communication for the
IEEE TRANSACTIONS ON COMMUNICATIONS. He has served as Guest Editor
of the special issue on Multimedia Network Radios in the IEEE JOURNAL
ON SELECTED AREAS IN COMMUNICATIONS, Associate Editor of the IEEE
TRANSACTION ON SIGNAL PROCESSING, and Membership Chair for of the
IEEE Information Theory Society. He is the recipient of the Meitec Junior
Fellowship Award from the Meitec Corporation in Japan, the George Van Ness
Lothrop Fellowship from the School of Engineering and Applied Science in
Princeton University, and the Research Initiation Award from the National
Science Foundation
\end{biography}

\end{document}